
\magnification=\magstep1

\def\dsp{\baselineskip=15pt plus 1pt minus 1pt}
\def\sles{\lower2pt\hbox{$\buildrel {\scriptstyle <}
   \over {\scriptstyle\sim}$}}
\def\sgreat{\lower2pt\hbox{$\buildrel {\scriptstyle >}
   \over {\scriptstyle\sim}$}}
\def\lapprox{\lower2pt\hbox{$\buildrel \lower2pt\hbox{${\scriptstyle<}$}
   \over {\scriptstyle\approx}$}}
\def\gapprox{\lower2pt\hbox{$\buildrel \lower2pt\hbox{${\scriptstyle>}$}
   \over {\scriptstyle\approx}$}}
\def\np{\vfill\eject}

\def\vb{\overline v}
\def\vbb{\overline{v^2}}
\def\s{\sigma}
\def\ss{\s^2}
\def\t{\tau}

\def\p{\partial}
\def\lx{\langle x^2 \rangle}
\def\lr{\langle r^2 \rangle}
\def\o{\over}
\def\ov{\overline}
\def\vi{v_i}
\def\vj{v_j}

\def\vib{\ov{v_i}}
\def\vjb{\ov{v_j}}
\def\vkb{\ov{v_k}}
\def\ui{u_i}
\def\uj{u_j}
\def\uk{u_k}
\def\uijb{\ov{\ui\uj}}
\def\ujkb{\ov{\uj\uk}}
\def\uxxb{\ov{u_1^2}}
\def\uyyb{\ov{u_2^2}}
\def\uzzb{\ov{u_3^2}}
\def\uxyb{\ov{u_1u_2}}
\def\l{\left}
\def\r{\right}
\def\vxb{\ov{v_1}}
\def\vyb{\ov{v_2}}
\def\vzb{\ov{v_3}}
\def\q{\qquad}
\def\k{\kappa}
\def\kk{\kappa^2}
\def\dvyb{\delta\vyb}
\def\duxyb{\delta\uxyb}
\def\ak{\alpha_K}
\def\a{\alpha}
\def\vsec{{\p^2\ov{v_2}\o\p x_1^2}}

\dsp
\centerline{\bf CAUSALITY IN STRONG SHEAR FLOWS}
\bigskip
\bigskip
\centerline{Ramesh Narayan, Abraham Loeb}
\medskip
\centerline{Harvard-Smithsonian Center for Astrophysics, 60 Garden St.,
Cambridge, MA 02138}
\medskip
\centerline{Pawan Kumar$^\dagger$}
\medskip
\centerline{Physics Department, Massachusetts Institute of Technology,
Cambridge, MA 02139}
\bigskip
\bigskip
\bigskip

\centerline{\bf ABSTRACT}
\medskip

It is well known that the standard transport equations violate
causality when gradients are large or when temporal variations are
rapid. We derive a modified set of transport equations that satisfy
causality. These equations are obtained from the underlying Boltzmann
equation.  We use a simple model for particle collisions which
enables us to derive moment equations non-perturbatively, i.e.
without making the usual assumption that the distribution function
deviates only slightly from its equilibrium value. We also retain
time derivatives of various moments and choose closure relations so
that the final set of equations are causal. We apply the model to two
problems: particle diffusion and viscous transport.  In both cases we
show that signals propagate at a finite speed and therefore that the
formalism obeys causality. When spatial gradients or temporal
variations are small, our theory for particle diffusion and viscous
flows reduces to the usual diffusion and Navier-Stokes equations
respectively. However, in the opposite limit of strong gradients the
theory produces causal results with finite transport fluxes, whereas
the standard theory gives results that are physically unacceptable.
We find that when the velocity gradient is large on the scale of a
mean free path, the viscous shear stress is suppressed relative to
the prediction of the standard diffusion approximation. The shear
stress reaches a maximum at a finite value of the shear amplitude and
then decreases as the velocity gradient increases.  The decrease of
the stress in the limit of large shear is qualitatively different
from the case of scalar particle diffusion where the diffusive flux
asymptotes to a constant value in the limit of large density
gradient. In the case of a steady Keplerian accretion disk with
hydrodynamic turbulent viscosity, the stress-limit translates to an
upper bound on the Shakura-Sunyaev $\alpha$-parameter, namely
$\alpha<0.07$.  The limit on $\alpha$ is much stronger in narrow
boundary layers where the velocity shear is larger than in a
Keplerian disk.


\vskip 2.0truecm
\noindent $^\dagger$ Alfred P. Sloan Research Fellow

\np
\centerline{\bf 1. INTRODUCTION}
\bigskip

According to the standard theory of diffusion the particle flux
increases linearly with the gradient of the particle density.  Thus,
the theory predicts that a sharp gradient will result in a divergent
flux. This violates causality, since the particle flux is obviously
limited by the finite particle speed. The standard theory runs into a
similar problem when the particle density changes on a very short
time scale locally (e.g. Morse \& Feshbach 1953, Narayan 1992).

A similar limitation is well-known in viscous interactions. In this
case the viscous stress tensor is linearly proportional to the
velocity gradient in the diffusion approximation. When this relation
is used in the Navier-Stokes equation one finds an instantaneous
propagation of viscous signals, in violation of causality. Since the
particle speed does not enter explicitly in the underlying theory,
the standard equations cannot be modified in a straightforward way
to limit the signal propagation speed.

The diffusion equation and the Navier-Stokes equation are valid only
when particles have suffered many collisions and their distribution
has relaxed to have weak spatial gradients and slow temporal
variations.  However, there are physical situations both in the
laboratory and in astrophysical gases where gradients are large on
the scale of a collisional mean free path or temporal changes are
rapid relative to the mean collision time.  Examples include
radiation hydrodynamics in optically thin media (Levermore and
Pomraning 1981), electron heat transport in laser produced plasma
(Max 1981), and viscous angular momentum transport in boundary layers
of accretion disks (Popham \& Narayan 1992).  One needs to go beyond
the standard fluid equations to model such conditions.

Specific prescriptions have been proposed to incorporate causality in
individual problems.  In radiation hydrodynamics, Levermore and
Pomraning (1981) introduced a flux-limiter such that, regardless of
the gradient, the radiative flux is never permitted to exceed the
product of the radiation energy density and the speed of light.  In
the accretion disk problem, Narayan (1992) proposed a modifying
factor for the viscosity coefficient, such that in steady state flows
the viscosity vanishes when the flow speed exceeds the maximum random
speed of the particles.

While the above prescriptions have worked well in their individual
applications, it would be useful to derive a general formalism that
automatically yields a causal limit to different transport phenomena.
We present such a formalism in this paper.  We base our work on the
Boltzmann equation which is strictly causal.  We make a simple
non-perturbative approximation to the scattering term in the
Boltzmann equation, take successive moments, and use appropriate
closure relations.  The standard diffusion approximation is then
recovered if we neglect certain terms involving time derivatives.
This approximation is valid if temporal variations are slow compared
to the collision time of particles, and spatial gradients are small,
but it breaks down whenever there are rapid variations and the flux
is limited by causality. When we retain all terms in the moment
equations, we obtain a causal set of equations.

The technical discussion in the paper is divided into two main
sections.  In \S2 we discuss the diffusion of particles in a fixed
background, while in \S3 we consider the stress tensor.

The diffusion problem provides a simple test-bed for our
approximations since it involves both temporal and spatial variations
of the gas properties.  We introduce the basic features of our
approach in \S2.1 where we discuss particle diffusion in one
dimension.  There are two main simplifications which permit us to
obtain a set of closed causal equations in this case.  First, we
write the scattering term of the Boltzmann equation in a simple
non-perturbative form. This explicit form allows us to take moments
of the Boltzmann equation and to obtain equations that are valid for
arbitrarily strong gradients in the particle density.  This
particular approximation of the scattering term is used throughout
the paper and is a key ingredient of our approach.  Second, we close
the equations by assuming that the second velocity moment of the
distribution function is constant.  In \S2.2 we explore the
properties of our one-dimensional diffusion equation and verify the
validity of these assumptions by comparing results with numerical
tests. We then generalize the theory to three-dimensional diffusion
in \S\S2.3 and 2.4.

Section 3 describes the viscous transport of momentum in the presence
of strong velocity gradients and presents the main results of this
paper.  In \S3.1 we derive a causal equation for the evolution of the
stress tensor.  We use the same approximation to the scattering term
that was employed in \S2.  However, since the elements of the stress
tensor are themselves second moments of the distribution function, we
close the equations at the level of the third moments rather than the
second moments.  We use the simplest approximation allowed, assuming
that all third moments vanish.  The causal equations we thus derive
reduce to the Navier-Stokes equation whenever the velocity gradients
are weak, but can also be used when the gradients are large.  We
apply the new equations to a number of problems involving a steady
state shear (\S3.2--\S3.5), and show that in the presence of a large
velocity gradient the viscous stress is suppressed compared to the
prediction of the standard diffusion approximation.  We consider the
effect of steady advection on the viscous shear stress in \S3.4 and
discuss bulk viscosity in \S3.5. In \S3.6 we extend our analysis to
rotating flows.  The presence of a Coriolis force introduces the
epicyclic frequency into the problem, and this leads to a
generalization of our formula for the modified shear stress.  In
\S3.7 we discuss the implications of this formula for accretion disks
and show that the $\alpha$ parameter introduced by Shakura and
Sunyaev (1973) has a strict upper limit whenever the viscosity is
mediated by hydrodynamic interactions.  Finally, we summarize the
main conclusions in \S4.

Parts of this paper overlap previous work in the subject.  The causal
particle diffusion equations which we derive in equations (2.1.11)
and (2.3.6) have appeared several times in the literature (Israel and
Stewart 1980 and references therein, Schweizer 1984).  Many of the
previous discussions have been somewhat phenomenological whereas we
derive the equations through a systematic procedure which reveals
clearly the specific assumptions which we make.  In particular, we
discuss the limits of the diffusion theory and identify exactly which
phenomena are described well by the theory and which are not.  Our
discussion of the shear problem, especially the effects of strong
shear and advection, appears to be largely new.  Some of the results
on rotating flows have been derived independently by Kato and Inagaki
(1993) whose preprint we received at a late stage of our work.

\np
\dsp
{\bf\item{2.} PARTICLE DIFFUSION}
\medskip
{\bf\item{2.1} The One-Dimensional Problem}
\medskip
We introduce our notation and explain our basic ideas and approximations by
discussing first a one-dimensional problem.  Consider a gas of light particles
diffusing in a fixed background of much heavier particles.
Let the light particles be described by a distribution function
$f(t,x,v)$ where $t$ is time, $x$ is particle position, and $v$ is
particle velocity.  We define various moments of $f$ in the usual way:
$$
n = \int fdv,~~~\vb = {1\over n}\int vfdv,~~~\vbb = {1\over n}\int v^2fdv.
\eqno{(2.1.1)}
$$
We assume that the particles experience no accelerations in between
scatterings.

The distribution function $f$ satisfies a Boltzmann equation of the form
$$
{\p f\over \p t} + {\p \over \p x}(vf) = \Gamma_{scatt} + \dot f_s,
\eqno{(2.1.2)}
$$
where $\Gamma_{scatt}$ describes the effect of scattering and $\dot
f_s (t,x,v)$ is the rate at which particles are introduced or removed
from a phase space element.  Let us write $\Gamma_{scatt}$ as
$$
\Gamma_{scatt} \equiv \Gamma^+ - \Gamma^- =
{1\over \t}(f_0 - f). \eqno{(2.1.3)}
$$
The term $\Gamma^-=-f/\t$ represents the rate at which particles are
removed from a phase space element, where $\tau (v)$ is the mean free
time which is in general a function of velocity.  $\Gamma^+=f_0/\t$
describes the phase space distribution of the particles after they are
scattered.  Provided we make $f_0$ a function of $f$ and normalize
$f_0$ so as to conserve particles, equation (2.1.3) will always be true.
Note that this equation does not imply that $f$ is close to $f_0$ in
any way.  Indeed many of the cases we consider in this paper
involve highly non-linear situations where $f$ is very different from
$f_0$.  In this respect we differ from the usual approach to transport
theory where one expands $f$ around an equilibrium $f_0$ and assumes
that the deviations are small (see e.g. Krook's equation [Shu 1992]).

In general, the scattering is a complicated function of velocity
which makes the Boltzmann equation very difficult to handle.  A major
simplification is achieved if we make two approximations (see
Appendix B of Grossman, Narayan, and Arnett 1993, also Kato and
Inagaki 1993).  First, we assume that $\tau$ is a constant,
independent of velocity.  Second, we make some simplifying
assumptions regarding the post-scattering distribution function
$f_0$, viz. we assume that each scattering is elastic in the frame of the
fixed background, that it conserves particles, and that it leads to
a total randomization of the initial velocities.  This allows us to write
simple
relations for the moments of $f_0$:
$$
\int f_0 dv = \int fdv=n,\qquad \int vf_0 dv = 0,\qquad \int v^2 f_0 dv =
n\vbb.\eqno{(2.1.4)}
$$
In the same spirit, we assume that the source function
$\dot f_s$ has zero mean velocity, and write
$$
\int \dot f_s dv = s(t,x),~~{1\over s}\int v\dot f_s dv = 0,~~
{1\over s}\int v^2 \dot f_s dv = \overline{v_s^2}. \eqno(2.1.5)
$$

With the above assumptions, we now take the first two moments of
equation (2.1.2).  Integrating equation (2.1.2) over $v$, we obtain
$$
{\p n\over \p t} + {\p \over \p x}(n\vb) = s.\eqno{(2.1.6)}
$$
Multiplying equation (2.1.2) by $v$ and integrating over $v$, we obtain
$$
{\p \over \p t}(n\vb) + {\p \over \p x} (n\vbb) = - {1\over \t} n\vb.
\eqno{(2.1.7)}
$$
Equations (2.1.6) and (2.1.7) are two coupled equations describing the
evolution of the particle moments.  Unfortunately, these equations involve
three different moments, $n,~ \vb $ and $\vbb$.  Therefore, we need a
closure relation.  We make the simplest assumption possible, namely that
the velocity dispersion of the particles, $\vbb$, is a constant, independent of
$x$ and $t$:
$$
\vbb \equiv \s ^2 = {\rm constant}. \eqno{(2.1.8)}
$$
This condition is exactly valid if all particles, including those created
by the source function $\dot f_s$, have a single speed, $v=\pm \s$,
and provided the scattering is elastic.  In a more general situation,
this condition may not be satisfied but one might hope that
it will perhaps still capture the
essential features of particle transport phenomena.
We discuss later the
degree to which the approximation (2.1.8) is valid
and where it breaks down.  Equation (2.1.7) now
becomes
$$
{\p \over \p t} (n\vb) + \s^2 {\p n\over \p x} = - {1\over \t} n\vb.
\eqno{(2.1.9)}
$$
Equations (2.1.6) and (2.1.9) provide a closed pair of equations for
the two moments $n$ and $\vb$.

If we ignore the time derivative in equation (2.1.9) we get the usual
formulation of diffusion, where the diffusive flux $n\vb$ is given by
$-D\p n/\p x$, with a diffusion constant $D = \s^2 \t$. In this
approximation $n$ satisfies the standard diffusion equation
$$
{\p n\over \p t} - D{\p^2 n\over \p x^2} = s, ~~D = \s^2 \t.\eqno(2.1.10)
$$
As is well-known, this equation violates causality (e.g. Morse \& Feshbach
1953, Narayan 1992).

If, on the other hand, we retain the time derivative in (2.1.9)
we obtain a modified diffusion equation which does preserve
causality.  Differentiating equation (2.1.9) with respect to $x$ and
combining with equation (2.1.6) we obtain
$$
{\p n\over \p t} - \s^2\t
\left( {\p^2 n \over \p x^2} - {1\over \s^2} {\p^2 n\over \p
t^2}\right) = s + \t {\p s \over \p t}. \eqno{(2.1.11)}
$$
This equation has been derived previously in the literature using the
theory of ``transient thermodynamics'' (Israel and Stewart 1980,
Schweizer 1984).  The most interesting feature of equation (2.1.11) is the
presence of the wave operator on the left hand side, which ensures
that signals cannot propagate faster than the {\it r.m.s.} particle
speed $\s$.

The Green's function $G_1(t, x)$ of the one-dimensional diffusion equation
(2.1.11) is obtained by solving the equation
$$
{\p G_1\over \p t} - \s^2\t \left( {\p^2 G_1\over \p x^2} - {1\over \s^2}
{\p^2 G_1\over \p t^2}\right) = \left[ \delta (t) + \t {d\over dt}
\delta(t)\right] \delta(x).\eqno{(2.1.12)}
$$
Morse and Feshbach (1953) have given the solution when the
right hand side consists only of the first term in the square brackets.
Modifying their solution for our case (Schweizer 1984, Nagel and
M\'esz\'aros 1985), we obtain
$$
 G_1(t, x)= \left [ {1\over 4\s \t}\left\{ I_0(u)+{t\over 2\t}
{I_1(u)\over u} \right\} + {1\over 2}\left\{ \delta(x-\s t)+\delta
(x+\s t) \right\} \right ] e^{-t/2\t},~ |x|\leq \s t,\eqno{(2.1.13)}
$$
$$
=0 \hskip 3in , |x| >\s t ,
$$
where $I_0$ and $I_1$ are modified Bessel functions and
$$
u = {1\over 2\t} \left( t^2 - {x^2\over \s^2}\right)^{1/2}. \eqno{(2.1.14)}
$$
Note that the Green's function cuts off for $|x|>\s t$.  This
demonstrates explicitly that equation (2.1.11) satisfies causality
with a maximum propagation speed of $\s$.

Equation (2.1.11) and the Green's function (2.1.13), which are
derived under several approximations, represent the physical
situation exactly in the particular case when all particles have the
same speed $\s$ and the same mean free time $\t$.  The $\delta$
functions and the causal fronts at $x = \pm \s \t$ in equation
(2.1.13) are of course a consequence of the mono-speed assumption.
The generalization of the Green's function for an arbitrary
distribution of speeds, $f(|v|)$, and an arbitrary dependence of the
mean free time, $\t (|v|)$, is straightforward and requires merely
integrating equation (2.1.13) over the particular
distributions\footnote*{The conditions, however, are different in
more complex gases, like plasmas.  Collisionless Landau damping may
dominate over collisional dissipation whenever the plasma properties
undergo strong temporal or spatial variations.}.

One effect that is not described by equations (2.1.11) and (2.1.13)
is the smoothing of density inhomogeneities by phase mixing (see e.g.
Binney \& Tremaine 1987).  Particles with different velocities travel
different distances within a collision period and therefore wash out
inhomogeneities.  The damping of inhomogeneities by free streaming of
particles is not described by equation (2.1.11) since it corresponds
to particles with a single speed.  However, for elastic collisions
this problem can be easily fixed by integrating the Green's function
(2.1.13) over the velocity distribution of particles.

Another way to improve the description of particle diffusion is to
drop the assumption of a constant $\ov{v^2}$ (which crudely speaking
corresponds to an isothermal system) in the moment equations.  We can
treat $\ov{v^2}$ as an independent moment and solve for it by
considering higher moments of the Boltzmann equation.  This extension
of the theory describes both particle and heat diffusion. We consider
up to third moments of the Boltzmann equation, and use a closure on
the fourth moments of the form $\ov{v^4}=\zeta(\ov{v^2})^2$ with
$\zeta$ equal to some constant (e.g. $\zeta =3$ for a Gaussian
distribution).  Let us generalize equations (2.1.4) and (2.1.5)
suitably for the higher moments,
$$
\int v^3f_0dv=0,\q \int v^4 f
dv=\zeta n(\vbb)^2,\eqno (2.1.15)
$$
$$
{1\o s}\int v^3\dot f_sdv=0 .
\eqno (2.1.16)
$$
We then obtain
$$ {\p\o\p t}(n\ov{v^2})+{\p\o\p
x}(n\ov{v^3})=s\ov{v_s^2},\eqno (2.1.17)
$$
$$ {\p\o\p
t}(n\ov{v^3})+\zeta{\p\o\p x}[n(\ov{v^2})^2]=-{1\o\t} n\ov{v^3}.\eqno
(2.1.18)
$$
By taking the $x-$derivative of equation (2.1.18) and
substituting equation (2.1.17) one gets a diffusion equation for
$n\ov{v^2}$,
$$
{\partial (n\ov{v^2})\over \partial t} -
\tau\left(\zeta{\partial^2\over \p x^2}[(n \ov{v^2})\ov{v^2}]-
{\p^2\o \p t^2}(n\ov{v^2})\right) =\tau{\p (s\ov{v_s^2})\o\p t}
+(s\ov{v_s^2}), \eqno (2.1.19)
$$
which is coupled to the diffusion
equation for $n$,
$$
{\p n\o\p t}- \tau\left({\p^2 [n\ov{v^2}]\o\p
x^2}-{\p^2n\o\p t^2}\right)= \tau{\p s\o\p t}+ s .  \eqno (2.1.20)
$$
Note that the limiting speed for the propagation of a pressure
perturbation is greater by a factor $\sim{\sqrt \zeta}$ than that of
a density perturbation.

Equations (2.1.19) and (2.1.20) provide a more accurate
representation of diffusion than the simpler version of the theory
presented earlier. Furthermore, these equations are causal as is
evident from the wave operator on the left side. In fact, these
equations also describe thermal diffusion since the quantity
$n\ov{v^3}$ in equations (2.1.17) and (2.1.18) represents the flux of
heat.  We do not discuss these extended equations further in this
paper but concentrate instead on the properties of the simpler
equation (2.1.11).

\bigskip
{\bf\item{2.2} Evaluating the Accuracy of the Approximate Theory}
\medskip

Figure 1 shows the Green's function $G_1$ at two different times,
$t = 0.1\t, {\rm and} ~10\t$.  For comparison, two other Green's functions
are also shown for each case.  One is the Green's function $G_s$ of the
standard diffusion equation (2.1.10).  The other is the Green's
function $G_M$ for a Maxwellian distribution of particles with an {\it r.m.s.}
one-dimensional velocity $\s$.  This is calculated by integrating equation
(2.1.13) over a Maxwellian in $\s$, and assuming $\t$ to be
independent of $v$.  At the detailed level of the shape of the
Green's function, neither equation  (2.1.10) nor equation (2.1.11) does a
particularly good job of fitting the Maxwellian Green's function,
particularly at early time.  This is not surprising since the two
theories have only one parameter $(D)$ and two parameters $(\t,
\s^2)$ respectively, and therefore they cannot describe the exact spatial
distribution of particles with a continuous velocity
distribution.

Although the shape of the Green's function $G_1$ differs significantly from
$G_M$, the spatial extent of $G_1$ is close to that of the Maxwellian Green's
function $G_M$ at all times (see figure 1). This
is in contrast to the Green's function $G_s$ of equation (2.1.10) which is
much too wide at early times (the usual signature of acausal
behavior). We now show that the mean square width $\lx$ of $G_1$ is,
in fact, exactly equal to $\lx$ of $G_M$.

Let us define $\lx$ for $G_1$ as follows:
$$ \lx = \int dx\, x^2G_1(t,x),~~{\rm where}~\int dx\,G_1(t,x)\equiv 1.
   \eqno{(2.2.1)} $$
Multiplying equation (2.1.12) by $x^2$ and integrating over $dx$ we find
$$
{\p \lx \over \p t} + \t {\p^2 \lx \over \p t^2} = 2\s^2\t, \eqno{(2.2.2)}
$$
whose solution with the appropriate boundary conditions is
$$
\lx = 2\s^2\t t+ 2\s^2\t^2\lbrack \exp (-t/\t)-1\rbrack.\eqno{(2.2.3)}
$$
At early times, i.e. for $t\ll\t$, we see that $\lx = \s^2t^2$ which
corresponds to particles streaming freely with a speed $\s$.  At late
times, however, we have $\lx \rightarrow 2\s^2\t t$ which corresponds
to the usual diffusion limit.

Since the Green's function $G_1$ corresponds exactly to the case of
particles with a single speed $\s$, equation (2.2.3) is the exact
result for the evolution of $\lx$ for a mono-speed population of
particles.  The interesting point is that the expression for $\lx$ is
directly proportional to $\s^2$.  Therefore, any distribution of
velocities which has a mean square velocity equal to $\s^2$ will
evolve exactly according to equation (2.2.3).  This equation is
therefore valid more generally than for just mono-speed particles ---
all it requires is that all the particles should have the same mean
free time.

If particles with different speeds do not have the same mean free
time, then obviously we do not have a perfect correspondence between
(2.2.3) and the exact result for $\lx$.  However, even in this case
we find that the theory performs quite well.  As a demonstration of
this result, suppose we have particles with a Maxwellian distribution of
one-dimensional speeds,
$$
f(v)dv = \sqrt {2\over \pi\s^2} {\rm exp} \left( - {v^2\over 2\s^2}\right)
dv, ~~ v\geq 0, \eqno{(2.2.4)}
$$
and let us assume that all the particles have the same mean free path $l$.
The mean free time is then a function of $v$,
$$
\t(v) = l/v. \eqno{(2.2.5)}
$$
To obtain the evolution of $\lx$ for this problem, we replace $\s^2$
by $v^2$ in equation (2.2.3) and integrate over the Maxwellian $f(v)dv$.  This
gives
$$
\lx = 2\sqrt {2\over \pi} \s lt + 2l^2\Bigl[ {\rm exp}\left({\s^2t^2\over
2l^2} \right) {\rm erfc}\left( {\s t\over {\sqrt 2}l}\right) - 1 \Bigr].
\eqno{(2.2.6)}
$$
At early and late times this has the following asymptotic
dependences: $\lx\rightarrow\ss t^2, ~t\ll\t$; $\lx\rightarrow
2(2/\pi)^{1/2}\s lt, t\gg\t$.  We can fit these dependences with
the single-speed result (2.2.3) provided we choose the mean free time
$\t$ to be
$$
\t = \sqrt {2\over \pi} {l\over \s}. \eqno{(2.2.7)}
$$
Figure 2 shows a comparison between the exact result (2.2.6) and the
approximate result (2.2.3) for this particular choice of $\t$.  The
agreement is very good.  In contrast, the standard diffusion
equation (2.2.10), which gives $\lx = 2\s^2\t t$ for all $t$, clearly makes a
large error for $t<\t$.  We expect our theory to give similar
good agreement with exact results for other distribution
functions $f(v)$ or prescriptions for the mean free time $\t (v)$,
provided we define the mean free time $\t$ appropriately.

We thus conclude that the causal diffusion theory developed here
provides a good representation of particle diffusion, and in
particular it predicts the mean square distance traveled by particles
very well at all times.

\bigskip
{\bf\item{2.3} Particle Diffusion in Three Dimensions}
\medskip

We work in a Cartesian representation with position indicated by the
coordinates $\vec r = (x_1,~ x_2,~ x_3)$ and velocity by $\vec v =
(v_1,~ v_2,~ v_3)$.  In the absence of any forces, but with scattering,
the Boltzmann equation gives
$$
{\p f \over \p t} + {\p \over \p x_i} (v_if) = \Gamma_{scatt} + \dot f_s
 = {1\over \t} (f_0 - f) + \dot f_s. \eqno{(2.3.1)}
$$
We use the summation convention on repeated indices.  Following
the one dimensional example of \S 2.1, we have written the scattering
term simply in terms of a mean free time $\t$.  We also make similar
assumptions as in equations (2.1.4) and (2.1.5), viz.
$$
\int f_0dv=n,\q\int v_if_0dv=0,\q\int\dot f_sdv=s,\q
{1\o s}\int v_i\dot f_sdv=0.
$$

Taking the zeroth and $v_j$th moments of (2.3.1) we obtain
$$
{\p n\over \p t} + {\p \over \p x_i}n \overline{v_i} = s, \eqno{(2.3.2)}
$$
$$
{\p \over \p t} (n\overline{v_j}) + {\p \over \p x_i}(n \overline{v_i v_j}) =
- {1\over \t} n\overline{v_j}. \eqno{(2.3.3)}
$$
These are not a closed set of equations because they involve the six
second moments $\overline{v_i v_j}$.  To close the equations, we
simplify the second moments by assuming that the
velocities are isotropically distributed and that the mean square velocity
is independent of $t$ and $\vec r:$
$$
\overline{v_iv_j} = {1\over 3} \ss\delta_{ij},\q\ss
= {\rm constant}. \eqno{(2.3.4)}
$$
This approximation is technically inconsistent, particularly at early
times.  While the mean square velocity itself can be constant,
for instance in the case of a mono-speed population of particles,
the assumption of isotropy is rather drastic and we may expect some
pathological behavior in the solutions in certain limits.  Nevertheless,
the resulting theory is simple and causal, and therefore we explore
it further.
Substituting equation (2.3.4) into equation (2.3.3) we find
$$
{\p \over \p t} (n\overline{v_j}) + {\s^2\over 3} {\p n\over \p x_j} =
- {1\over \t} n\overline{v_j}.\eqno{(2.3.5)}
$$
Combining equations (2.3.2) and (2.3.5) we then obtain
$$
{\p n\over \p t} -  {1\over 3} \s^2\t \left( \bigtriangledown ^2n -
{3\over \s^2}
{\p^2n\over \p t^2}\right) = s + \t {\p s\over \p t}. \eqno{(2.3.6)}
$$
As before, we find a wave operator appearing in the equation which
enforces causality.  Rather surprisingly, the relevant wave speed
appears to be $\s/\sqrt 3$.  This is however somewhat deceptive and
the true speed turns out to be $\s$ in certain situations as we show
in \S 2.4.  For reference we note that the standard diffusion equation in
three dimensions is
$$
{\p n\over \p t} - D \bigtriangledown^2n = s, \qquad D = {1\over 3}
\s^2 \t. \eqno{(2.3.7)}
$$

Morse and Feshbach (1953) show that the three-dimensional Green's function
$G_3(t, r)$ of the operator in equation (2.3.6) can be derived from
$G_1(t, x)$ discussed in \S 2.1.  The relation is
$$
G_3(t, r) = -{1\over 2\pi r} {d\over dr} G_1 (t, r),\eqno{(2.3.8)}
$$
with $\s$ in equation (2.1.13) replaced by $\s/\sqrt 3$ because of the modified
wave operator in equation (2.3.6).
The appearance of derivatives of the $\delta$-function in $G_3$ means
that this Green's function can give unphysical negative particle
densities under certain circumstances.  This may be somewhat
surprising considering the fact that the one dimensional Green's function was
not unphysical in any way.  There is, however, a significant physical
difference between the one-dimensional and three-dimensional
problems.  In one dimension, we showed that equation (2.1.11)
corresponds to a particular real physical situation, namely the case
of mono-speed particles.  This is unfortunately no longer true in
three dimensions.  There is no physical three dimensional system which
behaves exactly as described by equation (2.3.6). In particular, a
mono-speed population of particles does {\it{not}} obey this
equation.  The reason is associated with the closure assumption that the
second moment tensor of the velocity is isotropic (eq. 2.3.4).
This assumption is obviously violated close to an impulsive point source
before collisions take place.

As a final comment we note that if we consider a particular case
when all the gradients are restricted to one direction, say the
$x_1$ axis, then the three-dimensional equation (2.3.6) reduces to
$$
{\p n\o\p t}-{1\o3}\ss\t\l({\p^2n\o\p x_1^2}-{3\o\ss}{\p^2n\o\p t^2}
\r)=s+\t{\p s\o\p t}.\eqno (2.3.11)
$$
This is identical to the one-dimensional equation (2.1.11) except
for the identification $\ss\rightarrow\ss/3$.  As we have seen
in \S2.1 and \S2.2, the one-dimensional problem behaves quite
reasonably (it has no negative densities),
and therefore the three-dimensional version derived
here should be well behaved when gradients are nearly uni-directional.

In \S3, we derive a causal equation of evolution for the shear stress
and consider a number of flows with gradients in the velocities.  By
analogy with the particle diffusion problem we expect well behaved
results in the shear flow problem when gradients are in the same
direction everywhere.  All the situations we consider in \S3 do have
gradients limited to a single direction.

\bigskip
{\bf\item{2.4} Accuracy of the Three-Dimensional Diffusion Model}
\medskip We have seen that the Green's function $G_3$ contains the
derivative of a $\delta$-function which yields an unphysical density
at early times. Thus, the solutions to equation (2.3.6) cannot be
accurate in detail.  Following the discussion in \S 2.2, we can
however ask whether equation (2.3.6) is accurate in some more limited
sense.  For instance, does it fit some spatial moment of the particle
distribution?  The answer is that equation (2.3.6) does indeed do an
excellent job of predicting the evolution of the mean square distance
$\lr$ traveled by particles.

To show this we set $s = \delta(t) \delta^3(\vec r)$ in equation (2.3.6),
multiply the
equation by $r^2$, and integrate over $4\pi r^2 dr$.  We find that
$$
{\p\lr\over \p t} + \t {\p^2\lr\over \p t^2} = 2\s^2\t,\eqno{(2.4.1)}
$$
whose solution for the given boundary conditions is (cf. eq. 2.2.3)
$$
\lr = 2\s^2\t t + 2\s^2\t^2 [\exp(-t/\t)-1].\eqno{(2.4.2)}
$$
This result shows that $\lr$ varies as $\s^2 t^2$ at early time
$(t\ll \t)$, corresponding to free streaming at a speed $\s$.  (This
is a little surprising since equation (2.3.6) has an apparent speed
limit of $\s/\sqrt 3$ in the wave operator.  The Green's function
$G_3(t,r)$ too cuts off for $r>\s t/\sqrt 3$.  Mathematically, it is
the derivative of the $\delta$-function in $G_3$ which leads to an
{\it r.m.s.} particle distance of $\s t$ even though the Green's
function cuts off at a smaller radius.)  At late time, $\lr$ has
the standard dependence due to diffusion, viz.  $\lr = 2\s^2\t t$.

Motivated by the early time behavior of $\lr$ in equation (2.4.2), let us
find the ensemble averaged
evolution of $\lr$ for a set of particles with a
mean square velocity $\ss$ which are
injected at the origin at a time $t = 0$.
The Brownian motion of each particle is described by
Langevin's equation (Reif 1965),
$$
{d {\vec{v}}\over dt}= -{{\vec{v}}\over \tau} + {\vec{F}}(t)  ,
\eqno(2.4.3)
$$
The first term on the right hand side represents a frictional force,
where $\tau$ here is the same as the mean free time in our theory and
is taken to be a constant as per our assumptions.  The second term,
$\vec F(t)$, is a rapidly fluctuating force with zero mean.  We assume
that this force is arranged so that the mean square velocity $\langle
v^2\rangle \equiv\s^2$ is independent of time (again as in our model
of the scattering).  Multiplying equation (2.4.3) by ${\vec{r}}$,
averaging over the ensemble of particles, and noting that
$\langle\vec r\cdot\vec F\rangle=0$, we get
$$
\langle{d\over dt}({\vec{r}}\cdot{\vec{v}})\rangle = \s^2-{1\over \tau}
\langle{\vec{r}}\cdot{\vec{v}}\rangle   ,
\eqno(2.4.4)
$$
With the identity
$\langle{\vec{r}}\cdot{\vec{v}}\rangle={1\over 2}(d\langle r^2\rangle)$,
equation (2.4.4) admits the solution,
$$
\lr = 2\s^2\t t + 2\s^2\t^2 [\exp(-t/\t)-1].\eqno{(2.4.5)}
$$
which is identical to equation (2.4.2).
This means that equation
(2.3.6) predicts the evolution of the mean square distance $\lr$ {\it
exactly} for any population of particles with mean square velocity
$\s^2$ and constant mean free time $\t$.  This result clarifies in
what sense equation (2.3.6) is a good representation of diffusion in three
dimensions --- it is not very good at predicting the detailed shape
of the density distribution but it does provide an exact fit of
the mean square distance diffused by particles.

As an extension, we ask what would happen if the mean free times of the
particles are not all equal.  As before we consider the particular case
when all particles have the same mean free path $l$, i.e.
$$ \t(v) = l/v,\eqno{(2.4.6)} $$
and where the distribution of speeds is Maxwellian in three dimensions:
$$ f(v)dv = {4\pi\over\s_{_M}^3}\left({3\over 2\pi}\right)^{3/2}v^2\exp\left(
-{3v^2\over 2\s_{_M}^2}\right)dv, ~~ v\geq 0.\eqno{(2.4.7)} $$
After replacing $\s$ by $v$, one can integrate equation
(2.4.2) over this $f(v)$ to find
$$
\lr = 2\sqrt{8\over 3\pi}\, \s_{_M} l\,t - 4l^2{d\over d a}\left[
{1\over\sqrt a}\exp (w^2){\rm{erfc}} (w)\right]_{a\rightarrow 1} -
2\,l^2,~w={\s_{_M} t\over l\sqrt{6a}}.\eqno{(2.4.8)}
$$
This result has the following asymptotic behaviors:
$\lr\rightarrow \ss_{_M} t^2$ for $t\ll\t$; and $\lr\rightarrow
2(8/3\pi)^{1/2} \s_{_M} l\,t$ for $t\gg\t$.
Comparing with equation (2.4.2) we see that we can obtain agreement in the
behavior of $\lr$ in the two limits if we choose the mean free time
$\t$ in equation (2.4.2) to be
$$
\t = \sqrt{8\over 3\pi} {l\over\s_{_M}}.\eqno{(2.4.9)}
$$
Figure 3 shows a comparison between the exact result (2.4.8) and the
approximate result (2.4.2) when $\t$ is defined as in (2.4.9).  The
agreement is excellent, showing that our diffusion theory
predicts $\lr$ very well even when the mean free time is not
constant. Also shown in figure 3 is the result from the standard
diffusion equation, viz. $\lr = 2(8/3 \pi)^{1/2}\s_{_M} l\,t$, which is
very inaccurate at early time.

\np
{\bf\item{3.} STRESS TENSOR}
\bigskip
{\bf\item{3.1} Evolution Equation for the Stress Tensor}
\medskip

Having introduced our approach through the relatively simple problem of
particles diffusing through a fixed background, we now consider the case
when particles scatter off one another.  We decompose the velocity of a
particle $v_i$ into a mean velocity $\vib$ plus a relative velocity
$\ui$:
$$
v_i=\vib+\ui.\eqno (3.1.1)
$$
Particles experience an acceleration $\vec a=(a_1,a_2,a_3)$,
where the $a_i$ may be functions of time, position, and in
general velocity as well.  We ignore the possibility of particles
being added or removed from the system.  We thus have the Boltzmann equation
$$
{\p f\o\p t}+{\p\o\p x_i}(\vi f)+{\p\o\p\vi}(a_if)=
\Gamma_{scatt}={1\o\t}(f_0-f).\eqno (3.1.2)
$$

We depart from the discussion of \S 2 in the properties we assume for
the scattering.  Because the particles scatter off one another, the
relevant frame in which we should specify the properties of $\Gamma_{scatt}$
is not some fixed background frame but rather the local rest frame
of the gas of particles.  Since each scattering event conserves momentum,
the mean velocity of the particles must be preserved.  Thus
$$
(\vib)_0\equiv {1\over n} \int\vi f_0d^3v=\vib,\q
(\ov{\ui})_0\equiv {1\over n} \int\ui f_0d^3v=0.\eqno (3.1.3)
$$
Furthermore, we specify the second moments of the post-scattering distribution
function $f_0$ to be
$$
(\uijb)_0\equiv {1\over n} \int\ui\uj f_0d^3v=
(1-\xi){\ss\o3}\delta_{ij},\q \ss=\ov{\ui\ui}=
\uxxb+\uyyb+\uzzb.\eqno (3.1.4)
$$
In essence we assume that the scattering completely isotropizes the
particle velocities so that immediately after scattering, (i) there
are no off-diagonal velocity correlations such as $(\uxyb)_0$, and
(ii) the mean square velocity is independent of direction, i.e.
$(\uxxb)_0=(\uyyb)_0=(\uzzb)_0$.  A new feature in equation (3.1.4) is the
introduction of the parameter $\xi$, which allows for the possibility of
inelastic scattering ($\xi=0$ corresponds to elastic scattering).
The point is that particles heat up when there
are stresses present, and in real gases some of this heat is lost from the
system by, for example, radiative processes.  We introduce $\xi$ to model this
kind of ``cooling'' which has to be modeled if
we are interested in considering steady state situations.

Let us take the zeroth and first moments of the Boltzmann equation
(3.1.2).  These give the standard continuity and momentum equations:
$$
{\p n\o\p t}+{\p\o\p x_i}(n\vib)=0,\eqno (3.1.5)
$$
$$
{\p\vjb\o\p t}+\vib{\p\vjb\o\p x_i}=\ov{a_j}-{1\o n}{\p\o\p x_i}
(n\uijb).\eqno (3.1.6)
$$
In equation (3.1.6), there are two kinds of acceleration, a mean acceleration
$\ov{a_j}$ due to the external forces and a contribution from the gradient
of the stress tensor $n\uijb$.  In order to be general
and to permit velocity-dependent external forces, we
Taylor expand the acceleration $a_j$ in the form
$$
a_j=\ov{a_j}+{\p a_j\o\p\vi}\ui,\eqno (3.1.7)
$$
where $\ov{a_j}$ is the mean force and the second term is the excess
force on a particle due to its relative velocity.  Equation (3.1.7)
is quite general and the only approximation made is that the
expansion in velocity has been truncated at the linear term.

We now derive equations for the evolution of the stress tensor.  To
do this it is convenient to switch from the distribution function
$f(t,\{x_i\},\{\vi\})$ to another ``relative'' distribution function
$f_R(t,\{x_i\},\{\ui\})$ (cf. Kato 1970, Grossman et al. 1993) which
is a function of the relative velocities $u_i$.  This distribution
function satisfies a Boltzmann equation
$$
{\p f_R\o\p t}+{\p\o\p x_i}[(\vib+\ui)f_R]+{\p\o\p\ui}(\dot\ui f_R)=
{1\o\t}(f_0-f_R),\eqno (3.1.8)
$$
where $\dot\ui$ is the rate of change of the relative velocity of a
particle as we follow it along its trajectory:
$$
\dot\ui =\dot\vi-\dot{\vib}=\ov{a_i}+{\p a_i\o\p\vj}\uj-{\p\vib\o\p t}
-(\vjb+\uj){\p\vib\o\p x_j}
$$
$$
={\p a_i\o\p\vj}\uj-{\p\vib\o\p x_j}\uj+{1\o n}{\p\o\p x_j}
(n\uijb).\eqno (3.1.9)
$$
We now substitute equation (3.1.9) into equation (3.1.8) and take the $u_ju_k$
moment to obtain the equations of evolution of the stress components,
$$
{\p\o\p t}(n\ujkb)+{\p\o\p x_i}(n\vib ~\ujkb)+{\p\o\p x_i}(n
\ov{\ui\uj\uk})=-\l ({\p\vjb\o\p x_i}-{\p a_j\o\p\vi}\r )n\ov{\uk\ui}
$$
$$
-\l ({\p\vkb\o\p x_i}-{\p a_k\o\p\vi}\r )n\ov{\uj\ui}
+{(1-\xi)\o 3\t}n\ss\delta_{jk}-{1\o\t}n\ujkb.\eqno (3.1.10)
$$
Equivalently the equations for second moments are
obtained by combining (3.1.5) and (3.1.10),
$$
{\p\o\p t}(\ujkb)+\vib{\p\o\p x_i}(\ujkb)=-{1\o n}{\p\o\p x_i}(n
\ov{\ui\uj\uk})-\l ({\p\vjb\o\p x_i}-{\p a_j\o\p\vi}\r )\ov{\uk\ui}
$$
$$
-\l ({\p\vkb\o\p x_i}-{\p a_k\o\p\vi}\r )\ov{\uj\ui}
+{(1-\xi)\o 3\t}\ss\delta_{jk}-{1\o\t}\ujkb.\eqno (3.1.11)
$$

As usual, the above equations involve higher order moments (in this
case the third moments $\ov{\ui\uj\uk}$) and we need additional
relations to close the set of equations.  We faced such requirements
for closures in the discussion on diffusion in \S 2.  There we needed
closures for the second moments and we made the simplest assumption
possible, viz. that the second moments
are constant and isotropic.  Since the third moments are odd in velocity,
the simplest assumption here is to set all the third moments to zero:
$$
\ov{\ui\uj\uk}=0.\eqno (3.1.12)
$$
This is a reasonable approximation whenever the gradients of the
second moments are small.  More precisely, we expect odd moments in
the velocity to be negligible whenever there is sufficient spatial symmetry
in the neighborhood of the fluid element under consideration. Note that
equation (3.1.12) does {\it not} require velocity gradients to be small.
Indeed much of what we discuss later in the paper deals with large velocity
gradients.  We do however make sure that there is sufficient ``right-left''
spatial symmetry in the model problems we solve so that equation
(3.1.12) will be valid.
In the Appendix we derive
a conditions on the second and third derivatives of the mean velocity
which are necessary to make relation (3.2.11) viable
for a steady shear flow.
We should mention that there are other possible
closures for the third moments.  For instance, one could
use a ``diffusion approximation'' to relate the
third moments to gradients of the second moments, e.g.
$\ov{u_1^3}\propto\uxxb\t{\p\uxxb/\p x_1}$,
and so on.  Relations like this are reasonable, but they ultimately
correspond to setting time derivatives to zero in the third moment
evolution equations and using a Gaussian-like closure on fourth
moments (see equation [2.1.18]). The problem with the neglect of time
derivatives is that it may lead to a violation of causality.  As an analogy,
note that in the particle diffusion problem, if we neglect the time
derivative in (2.1.9) then we obtain the usual diffusion
approximation which gives the acausal equation (2.1.10).  In the same way
we expect a diffusion-like closure relation on third moments
to lead to a violation of causality.

Equations (3.1.5), (3.1.6) and (3.1.11) with third moments set to
zero provide 10 equations (one continuity equation, three momentum
equations, and six stress equations) that describe the evolution of
the 10 moments: $n, ~\vib$, and $\ov{u_iu_j}$. The Navier-Stokes
equation is a special case of these equations. When the spatial and
temporal derivatives are small on a collision scale
(3.1.11) gives the standard linear relation between the stress and
the velocity gradient.  This result can in turn be substituted into
equation (3.1.6) to get the Navier-Stokes equation.  While the
Navier-Stokes equation does not satisfy causality, our full equations
can handle strong velocity gradients and satisfy causality. The
combined equations (3.1.6) and (3.1.11) can potentially replace the
Navier-Stokes equation in situations where the gradients or the
temporal variations are arbitrarily large.  In the following sections
we explore the causal properties of these equations especially for
the particular case of steady flows with a one-dimensional velocity
gradient.  More work, however, is needed in order to examine the full
regime of applicability of equations (3.1.5), (3.1.6) and (3.1.11)
under general conditions.  Note that, at the level of our
approximation there is no independent equation of state.  Our system
of particles behaves essentially like an ideal gas, and whatever
prescription we may use for the cooling will determine the degree to
which the gas is non-adiabatic.  Deviations from an ideal gas
behavior could be introduced by allowing for additional degrees of
freedom in the particles but we do not explore this possibility here.

It is useful to write the stress tensor explicitly as the sum of the
diagonal terms and the off-diagonal terms.  The diagonal terms
can be decomposed into two parts:
$$ n\ov{u_i^2}=p+n\left(\ov{u_i^2}-{1\o 3}\ss\right),
\q p\equiv {n\ss\o 3},\eqno (3.1.13) $$
where the first term is the isotropic pressure.  The second term
describes the degree of anisotropy in the diagonal elements and
gives rise to bulk viscosity.  The off-diagonal
components of the stress tensor constitute the shear stress.  We
discuss the physics of the shear and bulk viscous stresses within our
formalism in the following sections by considering various
limiting cases of equation (3.1.11).

Before we proceed to discuss steady state conditions let us show
that the stress equation (3.1.11) satisfies causality.
For simplicity, consider a flow, in the absence of external forces,
where the mean velocity
at each point is oriented along the $x_2$ axis and its magnitude
$\ov{v_2}$ is a function of $x_1$. Since this flow is divergence-free
we take the particle density $n$ to be constant.  In addition we take
$\uxxb \approx \sigma^2/3$ to be constant, which is a valid assumption
for a sufficiently weak shear ($\ov{u_1u_2}\ll\sigma^2$).
Under these conditions the (1,2) component of equation (3.1.11)
together with the $x_2$-component of the momentum equation (3.1.6) yield,
$$ {\p \ov{v_2}\over \p t}-\nu_0\left({\p^2 \ov{v_2}\over\p x_1^2}
- {3\over \s^2}{\p^2\ov{v_2}\over \p t^2}\right)=0  , \eqno (3.1.14) $$
where $\nu_0=\sigma^2\t/3$ is the viscosity coefficient.
Equation (3.1.14) has the same structure and shares the same causal
properties as the diffusion equation (2.3.11). Thus, based
on the discussion in \S 2 we expect our shear flow theory
to be causal and well behaved in one dimension.

\bigskip
{\bf\item{3.2} Steady State Linear Shear}
\medskip

We begin our discussion of steady state shear flows by considering
a flow where there are no external accelerations, i.e.
$a_i=0$, and where the mean velocity at each point is oriented
along the $x_2$-axis.  We assume that the magnitude of the velocity
is independent of $x_2$ and $x_3$, and varies linearly with $x_1$, i.e.
$$
\vxb=0,\q \vyb=2Ax_1,\q \vzb=0.\eqno (3.2.1)
$$
We assume that the system is in a steady state,
and take $n$ to be a constant, as is consistent with the divergence-free
nature of the flow.  Under these conditions, the momentum equation (3.1.6)
shows that the various second moments have
no spatial gradients.  Furthermore, by the symmetries of the problem, the
only non-vanishing second moments are $\uxxb, ~\uyyb, ~\uzzb,~
\uxyb$. Thus, the four non-vanishing components of the
second moment equation (3.1.11) are given by
$$ (1-\xi )\ss-3\uxxb =0,\eqno (3.2.2) $$
$$ (1-\xi )\ss-3\uyyb - 12A\t\uxyb =0,\eqno (3.2.3) $$
$$ (1-\xi )\ss-3\uzzb =0,\eqno (3.2.4) $$
$$ 2A\t\uxxb+\uxyb =0.\eqno (3.2.5) $$
We add to these four equations the definition of $\ss$ given
in equation (3.1.4).

If we sum equations (3.2.2)---(3.2.4) we obtain the following relation,
$$
-\uxyb\cdot 2A={\xi\o\t}\cdot{1\o 2}\ss.\eqno (3.2.6)
$$
The left-hand side of this equation is the work done
(stress$\times$shear) per particle per unit time by the shear stress.
The right hand side represents the rate at which kinetic energy is
lost per particle as a result of cooling.  These two quantities
should obviously be equal in a steady state and equation (3.2.6)
shows that our system of equations is physically consistent.  More
importantly, it reveals the need for the cooling parameter $\xi$,
since without its presence there are no steady state solutions to the
equations (except for the trivial case of a zero shear).

Going back to equations (3.2.2)--(3.2.5), we can solve for $\xi$ to obtain
$$
\xi={8A^2\t^2\o 3+8A^2\t^2},\eqno (3.2.7)
$$
and from this we solve for all the second moments in
terms of $\ss$:
$$
\uxxb =\uzzb ={1\o 3+8A^2\t^2}\ss,\eqno (3.2.8)
$$
$$
\uyyb ={1+8A^2\t^2\o 3+8A^2\t^2}\ss,\eqno (3.2.9)
$$
$$
\uxyb =-{2A\t\o3+8A^2\t^2}\ss.\eqno (3.2.10)
$$
Note that in deriving these results we did not restrict the linear
shear parameter $2A$ in any way and therefore these formulae are
valid even for a large linear shear as long as the spatial derivatives
of the shear are sufficiently small (see the Appendix).
However, we did assume a
steady state.  Therefore, the results may not be applicable when the
flow varies on a collision time. Also, we assumed that all particles
have the same mean free time. Even when this is not the case, we
expect that we can choose $\t$ appropriately, so that our equations
will describe the behavior of the system well. This point was
demonstrated for the particle diffusion problem in \S2.2 and \S2.4,
and will be discussed further in the next subsection. Similarly,
there is no problem in principle with spatial variations of $\t$.
Finally, we note that the stress tensor depends on the form of the
cooling function assumed. However, the main qualitative features of
the results, such as the decrease in the stress $\ov{u_1u_2}$ at
large shear amplitude, appear to be universal.

Equations (3.2.7)--(3.2.10) show that the nature of the steady state
stress tensor depends on the magnitude of the dimensionless parameter
$A\t$.  In the limit of a weak shear, $A\t\ll 1$, we find that the\
velocity distribution of the particles is essentially isotropic,
i.e. $\uxxb\approx\uyyb\approx\uzzb\approx\ss /3$, and the shear stress
is given by
$$
n\uxyb =-n{\ss\t\o 3}\cdot 2A=-n\nu_0\cdot {\p\vyb\o\p x_1}.\eqno (3.2.11)
$$
This is the standard linear relation between
the viscous stress and the velocity shear.  The kinematic coefficient
of viscosity is
$$
\nu_0={\ss\t\o 3}.\eqno (3.2.12)
$$
The Navier-Stokes equation is based on a relation of the form (3.2.11)
for the shear stress.  This relation, plus an equivalent one for
the bulk viscosity which we discuss in \S 3.5, are introduced into
the momentum equation to give a closed set of dynamical equations
for the flow.  We see that this approximation
is valid only when the shear is weak,
i.e. only when the shear frequency $2A$ is small compared to the
scattering frequency $1/\t$.  For larger velocity gradients we
have to use the more complete set of equations presented here.

When the shear is strong, i.e. $A\t\gg 1$, we find a completely different
behavior than the one expressed by equation (3.2.11).  As
equations (3.2.8)--(3.2.10) show
the velocity distribution becomes highly
elongated along the $u_2$ axis, and the shear stress now is
{\it inversely} proportional to the velocity shear,
$$
\uxyb \approx -{1\o 4A\t}\ss.\eqno (3.2.13)
$$
The concept of a viscosity coefficient is not useful
in this regime, but if we were to formally define $\nu$ as in
equation (3.2.11) we would find that $\nu$ varies inversely as the square
of the shear.

The turnover and reduction of the shear stress when the shear is large
is a manifestation of causality in the theory.  From the expression given
in (3.2.10) we find that the maximum value of $|\uxyb |$ is
$$
|\uxyb |_{max}={1\o 2\sqrt{6}}\ss,\eqno (3.2.14)
$$
which shows that the shear stress $n\uxyb$ is always limited to be
smaller than the pressure ($p=n\ss /3$), regardless of the magnitude
of the shear.  This automatic stress limiter is a very natural and
welcome feature of the theory and is an improvement over the simple
relation (3.2.11).  The reason for the reduction of $\uxyb$ at a large
shear is not difficult to understand.
For large values of shear, particles coming from a mean free distance,
$u_1\t$, along the $x_1$ axis, develop a transverse velocity
difference $u_2\sim A (u_1\tau)$. Thus for $A\tau\gg1$,
$\uxxb\approx \uyyb/(A\t)^2 \sim \s^2/(A\t)^2$.
Therefore the stress, $|\ov{u_1u_2}|\sim \ov{u_1^2}A\tau\sim \s^2/A\tau$,
decreases for a large shear.

The asymptotic behavior of the shear stress in the limit of large
shear is qualitatively very different from that exhibited by the
diffusive flux in particle diffusion.  As we have shown above, the
shear stress actually decreases with increasing shear and goes to zero in
the limit $2A\rightarrow\infty$.  In particle diffusion on the other
hand, as the density gradient increases the diffusive flux asymptotes
to a constant value which is equal to the product of the particle
density and the {\it r.m.s.} particle speed (Levermore and Pomraning
1981, Narayan 1992).

\bigskip
{\bf\item{3.3} Shear Stress from Numerical Simulations}
\medskip

The solution written down in equations (3.2.7)--(3.2.10) is valid even for
highly non-linear steady state shears, where by non-linear we mean
that $f$ deviates considerably  from $f_0$.  We have tested this
solution using numerical simulations and describe some of our results
here.

We set up a shearing cell extending from $x_1=-0.5$ to $+0.5$ with a
large number $N$ of particles and with boundary conditions imposed so
as to enforce the mean velocity to follow equation (3.2.1).  The particles
scatter with a specified mean free time $\t$ and are re-injected with
the appropriate cooling $\xi$ and an isotropic Maxwellian velocity
distribution in the local rest frame. We let the system evolve until
it achieves steady
state and then compare the results with the theoretical predictions.
Figure 4 shows a case with $N=10^5, ~2A=1, ~\t =1, ~(1-\xi )\ss =1$.
We display the variation of the moments with time.  Note that the
system achieves a steady state in less than 10 mean free times.

Figure 5 shows the distribution of the velocity components $u_1$ and
$u_2$ at the end of the run for the three cases that we have
simulated.  The steady state distribution $f$ is only
slightly deformed from the Maxwellian post-scattering distribution
$f_0$ when $2A\t=0.1$.  This is a case of weak shear and it is
understandable that $f$ is only slightly perturbed from $f_0$.
Indeed perturbation theory is valid in this case, and so is the
standard weak shear formula equation (3.2.11) for the shear stress.  In the
other two cases that we have presented, viz. $2A\t=1,~10$, we find
that $f$ is strongly distorted by the shear.  We are therefore definitely
into the regime of non-linear shear.
Nevertheless, we have verified that all the second moments
calculated analytically (eqs. [3.2.8]--[3.2.10]) are in excellent
agreement with the numerical simulation.

We next consider a different situation where the particles have a
constant mean free path $l$ rather than a constant $\tau$.  Figure 6
shows $\ov{u_1u_2}/\s^2$ as a function of the shear
amplitude $2Al/\sigma$ in this case. The points are results
from a simulation of $N=10^3$ particles, and the curves
were obtained from equation (3.2.10) when we set $\tau=l/\sigma$
(dashed line) or $\tau=0.55l/\sigma$ (solid line).
Note the interesting result that equation (3.2.10) predicts the
maximum value of stress accurately for particles with a constant $l$,
although it was derived for the constant $\tau$ case.

\bigskip
{\bf\item{3.4} Steady State Shear with Advection}
\medskip

The stress-limiter discussed in \S 3.2 provides a strong
indication that our
theory satisfies causality.  To demonstrate causality in a different
and possibly more transparent way we consider here a generalization of
the previous case where, in addition to the shear in $\vyb$, we now
also allow motion along the $x_1$ axis ($\vxb$). We refer to this additional
motion as {\it advection}.  We do not restrict $\vxb$ or $\vyb$ in any way
except to assume that the gradients of both velocities are in the
$x_1$ direction and that there are no external forces.
Also, we retain the assumption of a steady state.
The reader may notice a close analogy between the flow
considered here and that in a steady state accretion disk where $x_1$
and $x_2$ would refer to the radial ($R$) and azimuthal ($\phi$)
directions respectively.  We do not, however, discuss rotation in this
section (but see \S3.6 \& 3.7).

In a steady state, the continuity equation (3.1.5), the $x_2$ component of the
momentum equation (3.1.6), and the $\uxyb$ component of the second moment
equation
(3.1.11) give
$$
{1\o n}{\p n\o\p x_1}=-{1\o\vxb}{\p\vxb\o\p x_1},\eqno (3.4.1)
$$
$$
\vxb{\p\vyb\o\p x_1}=-{\p\uxyb\o\p x_1}-{\uxyb\o n}{\p n\o\p x_1} =
-{\p\uxyb\o\p x_1}+{\uxyb\o\vxb}{\p\vxb\o\p x_1}.\eqno (3.4.2)
$$
$$
\vxb{\p\uxyb\o\p x_1}=-\uxxb{\p\vyb\o\p x_1}-{1\o\t}\uxyb - \uxyb{\p\vxb
\over\p x_1} .\eqno (3.4.3)
$$
Substituting (3.4.2) into (3.4.3) and rearranging terms we find
the following expression for the steady state shear stress $\uxyb$:
$$
\uxyb=-\l({1\o 1+2\t\p\vxb/\p x_1}\r)\l(1-{\vxb^2\o\uxxb}\r)
\uxxb\t{\p\vyb\o\p x_1}.\eqno (3.4.4)
$$
This relation has several interesting features.

First notice that if $\vxb=0$ the viscosity coefficient
that relates $\uxyb$ to the shear $\p\vyb/\p x_1$ is just $\uxxb\t$.
This agrees exactly with the results of the previous analysis (eqs.
[3.2.10] and [3.2.8]).  When there is advection, the viscosity coefficient
is modified by two factors.  Consider first the factor
$(1-\vxb^2/\uxxb)$, which reduces the viscosity whenever there is a
steady flow, and is very similar to results previously obtained by
Narayan (1992).  To interpret this result, we need first to ask under
what conditions we can have a shear in a steady flow such as we are
considering.  Since we have ignored external forces, there is no
outside mechanism that can induce shear in the flow; the
only reason for there to be a shear is a downstream boundary
condition.  For instance, let us suppose that the flow comes in from
negative $x_1$ with a positive $\vxb$ and with $\vyb=0$.  Let us
assume that at some large positive $x_1$ the flow meets a sideward
moving interface of some sort --- a ``conveyer belt'' in the language
of Syer and Narayan (1993).  The downstream boundary condition can be
satisfied only if the flow sets up a shear $\p\vyb/\p x_1$.  Now,
information about this downstream requirement has to be transported
upstream by the particles themselves.  As the flow velocity (as
measured in the frame in which $\p/\p t=0$) increases, the net flux of
upstream moving particles reduces because only a fraction of the
particles has large enough negative velocities to overcome the
advection.  Therefore, the number of particles that participate in the
shear stress reduces and this is reflected by a reduction in the
viscosity coefficient

Equation (3.4.4) shows that the viscosity coefficient goes to zero
when $\vxb=(\uxxb)^{1/2}$.  In particular, when the shear is weak, we
know that $\uxxb\approx\ss/3$, and the cutoff occurs at
$\vxb=\s/\sqrt{3}$.  The quantity $(\uxxb)^{1/2}$ is the limiting
speed of particles in the $x_1$ direction in our theory.  When $\vxb$
is equal to this speed, no particles are able to move fast enough to
beat the advection and as a result no viscous stress is transmitted
upstream.  This is the reason for the cut-off of the viscosity
coefficient. In real gases the cut-off may be more gradual due to the
existence of a high-velocity tail in the particle distribution function.

We note an apparent paradox in the fact that the viscosity
coefficient in equation (3.4.4) actually changes sign when
$\vxb^2>\uxxb$, which means that the shear stress begins to point in
the same direction as the velocity shear.  It implies that the work
done by the shear stress, $-\uxyb\p\vyb/\p x_1$ (see equation 3.2.6),
is negative, i.e. the shear stress extracts heat energy from the flow
and converts it into mechanical energy.  Since this is not
reasonable, we interpret the change of sign in (3.4.4) to mean that,
when $\vxb^2>\uxxb$ there can be no steady state shear in the flow.

We now come to the second factor $(1+2\t\p\vxb/\p x_1)^{-1}$ in equation
(3.4.4).  It shows that any divergence in the advection velocity
modifies the viscosity coefficient.  This is fairly straightforward to
interpret.  Consider first the case when $\p\vxb/\p x_1>0$, which
corresponds to an expansion of the flow.  In this case, as
a particle moves from one region of the flow to another it finds that
its velocity becomes closer and closer to the local bulk velocity
$\vxb$.  Therefore, the distance that a particle is able to move in
``comoving fluid coordinates'' before it is scattered is reduced.  As
a result, the number of particles that can reach any point in the flow
from downstream is reduced, and this causes a reduction in the shear
stress.  The case of compression, when $\p\vxb/\p x_1<0$, is similar
up to a point.  In the presence of compression, downstream particles
find it easier to move upstream and this enhances the shear stress.
The interesting point is that the shear stress diverges when
$2\t\p\vxb/\p x_1=-1$.  What this means is that a flow with such a
large compression is unphysical.  When $2\p\vxb/\p x_1<-1/\t$, the flow
compresses down to a zero volume in less than one scattering time.
Obviously such a situation can occur only in transient flows and not
in a steady state.

Equation (3.4.4) is a general result for steady flows which clarifies
several physical issues.  However, because the expression for the
stress is given in terms of $\uxxb$, whose magnitude relative to $\ss$
depends on the strength of the shear, it does not
provide a useful formula for the viscosity coefficient.  To find a
more practical formula we restrict ourselves to a steady flow where $\vxb$,
$n$ and $\s^2$ are constant. In this situation the momentum equation gives
$$ {\p\uxyb\o\p x_1} =-\vxb{\p\vyb\o\p x_1}.\eqno (3.4.5) $$
and the four relevant components of the second moment equation (3.1.11) are
$$ \vxb{\p\uxxb\o\p x_1}+{\uxxb\o \t} -{(1-\xi )\o 3\t}\ss=0,\eqno (3.4.6) $$
$$ -{(1-\xi )\o 3\t}\ss +\vxb{\p\uyyb\o\p x_1}+{\uyyb\o \t}
 +2\uxyb {\p\vyb\o\p x_1} =0,\eqno (3.4.7) $$
$$
-{(1-\xi )\o 3\t}\ss +\vxb{\p\uzzb\o\p x_1}+{\uzzb\o\t} =0,\eqno (3.4.8)
$$
$$
\uxxb{\p\vyb\o\p x_1}+\vxb{\p\uxyb\o\p x_1}+{\uxyb\o\t} =0.\eqno (3.4.9)
$$

 From the momentum equation (3.1.6) we
find that $\p\uxxb/\p x_1$=0, and since
$\ss$ is independent of position, it follows that
$\p\uyyb /\p x_1 = -\p\uzzb /\p x_1$.  We
now solve equations (3.4.5)--(3.4.9) making use of these relations,
and find
$$
\uxyb =-{\ss\t\o 3+8A^2\t^2}\l (1-{3\vxb^2\o\ss}\r )
{\p\vyb\o\p x_1},\eqno (3.4.11)
$$
which is just equation (3.2.10) modified by the causality factor
$(1-3\vxb^2/\ss )$.  This provides a useful special case of
equation (3.4.4) for flows with uniform advection.

\bigskip
{\bf\item{3.5} Bulk Viscosity}
\medskip

We now explore the nature of bulk viscosity in our theory.
We consider a flow where the mean velocity at each point is
oriented along the $x_1$-axis and where all gradients are also
along $x_1$.  Let us define the total time derivative $d/dt$ by
$$
{d\o dt}={\p\o\p t}+\vxb{\p\o\p x_1}.\eqno (3.5.1)
$$
The continuity and momentum equations (ignoring external accelerations)
give
$$
{dn\o dt}+n{\p\vxb\o\p x_1}=0,\eqno (3.5.2)
$$
$$
n{d\vxb\o dt}+{\p p\o\p x_1}+{\p\o\p x_1}\l [n\l (\uxxb-{\ss\o 3}
\r )\r ]=0,\eqno (3.5.3)
$$
where in (3.5.3) we have split the stress tensor into the sum of the
pressure $p$ and an anisotropic term which will be associated
with bulk viscosity (cf. eq. [3.1.13]).

By the symmetry of this problem, there are only three non-vanishing
second moments, namely $\uxxb, ~\uyyb, ~\uzzb$.  Ignoring the
acceleration terms, these components evolve according to
$$
{d\uxxb\o dt}+\l (2{\p\vxb\o\p x_1}+{1\o \t}\r )\uxxb
-{1-\xi\o 3\t}\ss =0,\eqno (3.5.4)
$$
$$ {d\uyyb\o dt}+{\uyyb\o \t} -{1-\xi\o 3\t}\ss=0,\eqno (3.5.5) $$
$$ {d\uzzb\o dt}+{\uzzb\o \t} -{1-\xi\o 3\t}\ss =0.\eqno (3.5.6) $$
Adding equations (3.5.4)--(3.5.6) we find
$$
-\uxxb\,{\p\vxb\o\p x_1}=\l ({d\o dt}+{\xi\o\t}\r )\cdot {1\o 2}
\ss.\eqno (3.5.7)
$$
This is analogous to equation (3.2.6).  The left-hand-side
gives the work done per particle per unit
time.  Note that the stress $n\uxxb$ is the sum of the pressure and
the bulk viscous contribution.  The right hand side of (3.5.7) is the
rate at which particles acquire kinetic energy, some of which is lost
through cooling.  Equation (3.5.7) thus expresses energy conservation.

Let us first consider the case when the velocity gradient $\p\vxb /\p x_1$
is small compared to $1/\t$.  In this case we expect $\uxxb\approx
\uyyb\approx\uzzb\approx\ss /3$.  Equation (3.5.4) gives
$$
{1\o\t}\l (\uxxb-{\ss\o 3}\r )=-2\uxxb {\p\vxb\o\p x_1}
-{d\uxxb\o dt}-{\xi\o 3\t}\ss.\eqno (3.5.8)
$$
Substituting $\uxxb\approx\ss /3$ into the right hand side and using
equation (3.5.7) we find
$$
n\l (\uxxb-{\ss\o 3}\r )\approx -{4\o 3}\nu_0n{\p\vxb\o\p x_1},\eqno (3.5.9)
$$
where $\nu_0$ is the kinematic coefficient of shear viscosity
introduced in equation (3.2.12).  We thus recover the standard result, viz.
that in the limit of weak velocity gradients, the shear and bulk
viscosity coefficients are related by a factor of $4/3$.  If this
relation and the analogous equation (3.2.11) are substituted in the momentum
equation we obtain the Navier-Stokes equation for the evolution of a
viscous fluid.  These relations are, however, valid only for weak
velocity gradients and for slow variations of the flow parameters.
When these conditions are violated we need
the more complete theory presented in this paper.

Let us now proceed to the case when the velocity gradient is not
necessarily small.  For simplicity we assume that the velocity
gradient is independent of $x_1$:
$$ {\p\vxb\o\p x_1}=2C=\ {\rm constant},\eqno (3.5.10) $$
Let us also assume that the temperature, which is proportional to $\ss$,
is the same everywhere and does not change with time.
It follows from equation (3.5.7) that $\uxxb$ must also be independent
of position and time. It is now straightforward to
solve equations (3.5.5)--(3.5.7), which yields
$$ \xi =-{4C\t\o 3+8C\t},\eqno (3.5.11) $$
$$
\uxxb ={1\o 1+8C\t /3}{\ss\o 3}.\eqno (3.5.12)
$$
Note that $\xi$ is negative for positive $C$.  This is because
particles cool when there is an expansion of the flow and we need a source
of heating (rather than cooling) to maintain the moments at a constant
value as we have assumed.

For a small velocity gradient, $|C\t|\ll 1$, equation (3.5.12) recovers
the usual result
$$
\uxxb-{\ss\o 3}\approx -{4\o 3}\cdot{\ss\t\o 3}\cdot 2C,\eqno (3.5.13)
$$
which is identical to equation (3.5.9).  When $C\t$ is large and positive we
find $\xi\rightarrow -1/2$ and $\uxxb/\ss\rightarrow 0$.  For an
extremely rapid expansion the particles are very cold in the $x_1$
direction and their velocity distribution is restricted nearly to the
$u_2$--$u_3$ plane.  This is reasonable.  When $C\t$ is large and
negative, equations (3.5.11), (3.5.12) appear to reveal a problem.  If $C\t
<-1/4$, then we find $\uxxb >\ss$ which is physically inconsistent.
There is, however, a simple explanation for this result.  Precisely
when $C\t=-1/4$ we
notice that $\xi=1$.  Going back to the original equation (3.1.4)
where $\xi$ was defined, we see that $\xi=1$ is a physical upper limit
to the cooling, representing the case when all the energy in a
particle is removed entirely in a collision.  However,
$C\tau=-1/4$ corresponds to an infinite compression of the gas in one
collision time, which generates
an infinite amount of heat. Clearly our cooling term is unable to remove this
heat, and thus the assumption of a steady state breaks down.
The present limit on $C\t$ is exactly equivalent to the
limit on $\p\vxb/\p x_1$ discussed in \S 3.4, and these have the
same underlying physics.

\bigskip
{\bf\item{3.6} Shear Viscosity in Differential Rotation}
\medskip
We consider finally a shearing rotating flow as an example of an
application with velocity-dependent accelerations.  This is also of
particular importance in astrophysics because of applications in
accretion disks.  Kato and Inagaki (1993) discuss rotating flows in
greater detail than we do, but they have limited themselves to a small
shear, whereas the following discussion is applicable even for a large
shear. (See also Goldreich \& Tremaine 1978, who have derived an
expression for the shear stress in a rotating disk of particles in
Saturn's ring using a different approach.)

We consider a rotating flow in a local Cartesian approximation, called
the shearing sheet approximation (Goldreich \& Lynden-Bell 1965,
Narayan, Goldreich \& Goodman 1987).  Let the flow take place in a
central potential which produces an inward radial acceleration $g(R)$.
Consider a reference radius $R_0$ and let $\Omega =
[g(R_0)/R_0]^{1/2}$ be the equilibrium angular velocity such that
centrifugal acceleration just cancels $g$ at $R=R_0$.  We work in a
frame which rotates with angular velocity $\Omega$ and is centered on
$R_0$.  We take our local Cartesian grid to have $x_1 = R-R_0,~x_2$
along the azimuthal direction, and $x_3$ parallel to $\vec \Omega$.
We have arranged so that a particle at rest at $x_1=x_2=x_3=0$
is in equilibrium.  However, particles at non-zero $x_1$ experience an
additional acceleration of the form $g' x_1$, arising from a
combination of the central potential and centrifugal force; we have
$g' = dg/dR+\Omega^2$.  In addition, moving particles in the
rotating frame experience a Coriolis force.  Therefore, the three
components of the acceleration are given by
$$
a_1=g' x_1+2\Omega v_2,\q a_2=-2\Omega v_1,\q a_3=0.\eqno{(3.6.1)}
$$
For a truly spherical central potential, $a_3\not=0$, but we have
simplified matters by assuming a cylindrically central
potential (as appropriate for a thin disk).

Let us consider a steady state shear flow in this rotating frame, and
let us assume that $n$ is constant and that (cf. \S 3.2)
$$
\vxb=0,\q \vyb=2Ax_1,\q \vzb=0.\eqno(3.6.2)
$$
The four non-vanishing second moments, $\uxxb, ~\uyyb,~\uzzb$, and $\overline
{u_1 u_2}$, satisfy the following equations
(compare to equations 3.2.2---3.2.5):
$$ 3\uxxb - (1-\xi)\ss - 12\Omega\tau\, \overline{u_1 u_2} =0,\eqno(3.6.3) $$
$$ 3\uyyb - (1-\xi)\ss + 12B\tau\, \overline{u_1 u_2} =0,\eqno(3.6.4) $$
$$ 3\uzzb - (1-\xi)\ss = 0,\eqno(3.6.5) $$
$$ 2B\tau\,\uxxb - 2\Omega\tau\,\uyyb + \overline{u_1 u_2} = 0,\eqno(3.6.6) $$
where we have defined the vorticity $2B$ by
$$
2B = 2A + 2\Omega.\eqno(3.6.7)
$$
As before, summing equations (3.6.3), (3.6.4) and (3.6.5) we obtain the usual
energy conservation relation
$$
-\overline{u_1 u_2}\cdot 2A = {\xi\o \tau}\cdot{1\o 2}\ss.\eqno(3.6.8)
$$
Solving the equations for the cooling constant $\xi$ and the various
second moments we find
$$
\xi = {8A^2\t^2\o 3 + 8A^2\t^2 + 12\kk\t^2},\eqno(3.6.9)
$$
$$
\uxxb = {1+8\Omega^2\t^2 + 2\kk\t^2\o 3 + 8A^2\t^2 + 12\kk\t^2}\ss,
\eqno(3.6.10)
$$
$$
\uyyb = {1+8B^2\t^2 + 2\kk\t^2\o 3 + 8A^2\t^2 + 12\kk\t^2}\ss,\eqno(3.6.11)
$$
$$
\uzzb = {1+4\kk\t^2\o 3 + 8A^2\t^2 + 12\kk\t^2}\ss,\eqno(3.6.12)
$$
$$
\uxyb = - {2A\t\o 3 + 8A^2\t^2 + 12\kk\t^2}\ss,\eqno(3.6.13)
$$
where $\kk$ is the epicyclic frequency,
$$
\kk = 4\Omega B.\eqno(3.6.14)
$$

These are generalizations of the results of \S 3.2 when there is
rotation.  We notice that the results now depend on two dimensionless
parameters, $A\t$ and $\k\t$, rather than one, because the flow now has two
frequencies associated with it, the shear $2A$ and the epicyclic
frequency $\k$.

We may consider various limits of equation (3.6.13).  If $A\t,~\k\t \ll 1$,
which corresponds to very rapid collisions, we recover the usual result
$$
\uxyb = -{\ss\t\o 3}\cdot 2A = -\nu_0{\p\vyb\o\p x_1},\eqno(3.6.15) $$
which is the standard formula for shear viscosity.  In this limit,
rotation plays no role.  Similarly, if $\k\t \ll 1$ and if $A\t \gg 1$,
i.e. if we have strong shear in the presence of weak rotation, we again
recover equation (3.2.10).  On the other hand, if $\k\t \gg 1$, then even if
the shear is weak (i.e. $A\t \ll 1$), the shear stress is suppressed
relative to the non-rotating case (see also Kato and Inagaki 1993),
$$
\uxyb \simeq - {1\o 1+4\kk\t^2}\cdot {\ss\t\o 3}\cdot 2A.\eqno(3.6.16)
$$
The reason for this is straightforward.  When $\kappa \gg 1/\t$ a
particle undergoes many epicycles within a mean free time and
travels a distance equal only to the radius of an
epicycle in the $x_1$ direction.  Therefore, the effective mean free
path is much shorter than that in the non-rotating case and this
suppresses the viscosity.  Finally, if $A\t, \k\t$ are both $\gg 1$,
then the viscous stress is reduced by an even larger factor.

In the derivation of the basic equations we had set the third moment of
the velocity to zero. Physically this condition is satisfied when
the velocity distribution of particles has a reflection symmetry,
or in particular when the distribution is constant on ellipses.
This turns out to be a very good approximation
for shear flows in rotating systems, where the
epicyclic motion of particles results
in a quite symmetric velocity distributions (see figure 8).

Obviously, the shear stress $n\uxyb$ in the presence of rotation is
limited by the two parameters discussed above, instead of just one as
in the non-rotating case (eq. [3.2.14]).  The maximum stress is thus
smaller than in non-rotating flows.  We discuss the stress limiter
further in \S 3.7, where we specialize some of these results to the
case of a Keplerian disk.

To demonstrate causality more explicitly, we restore the temporal and spatial
derivatives in the $\vyb$ component of the momentum equation and the
equation for $\uxyb$:
$$
{\p\vyb\o \p t} = - {\p\o \p x_1}(\uxyb),\eqno(3.6.17)
$$
$$
{\p \ov{u_1 u_2}\o \p t} +\uxxb{\p\vyb\o\p x_1}+(\uxxb-\uyyb)2\Omega
=-{1\o\t}\uxyb.\eqno(3.6.18)
$$
We continue to assume that $n$ is constant and that $\vxb = 0$.
This assumption is technically inconsistent since variations in
$\vyb$ will cause fluctuations in $\vxb$
(through the Coriolis acceleration) which will lead to compressive
motions and changes in $n$.  However, compression in $n$ will cause
sound waves which will complicate the analysis.  We filter out the sound
waves by ignoring variations in $\vxb$ and concentrating only on
$\vyb$ and $\uxyb$.  Let us further assume that the velocity shear $\p\vyb/
\p x_1$ is small.  In this limit, we have $2A\t\ll 1$, $B\approx
\Omega$, $\kappa\approx 2\Omega$.
Equations (3.6.10) and (3.6.11) then indicate that
$\uxxb\approx\uyyb\approx\ss/3$.  Substituting these relations into
(3.6.18), we find
$$
{\p \ov{u_1 u_2}\o \p t} +{\ss\o 3}{\p\vyb\o\p x_1}
=-{1\o\t}\uxyb.\eqno(3.6.19)
$$

We now consider small perturbations in $\vyb$ and $\uxyb$:
$$
\vyb\rightarrow\vyb+\dvyb,\q \uxyb\rightarrow\uxyb+\duxyb.\eqno (3.6.20)
$$
Equations (3.6.17) and (3.6.19) then combine to give
$$
{\p\dvyb\o \p t} - {\ss\t\o 3}\left[{\p^2\dvyb\o\p x^2}
-{3\o \ss} {\p^2\dvyb\o \p t^2}\right] = 0\eqno(3.6.19)
$$
This is analogous to equation (2.3.11) for particle diffusion and has the
familiar wave operator which enforces causality.  Thus, we have proved that
even in the presence of strong rotation, viscous signals travel at
a finite speed and satisfy causality.

\bigskip
{\bf\item{3.7} Limits on $\alpha$--Viscosity}
\medskip
Shakura and Sunyaev (1973) used simple scaling arguments to write the kinematic
viscosity coefficient in a turbulent thin accretion disk in the form
$$
\nu = \alpha {c_s^2\o \Omega_K},\q\a\leq 1,\eqno(3.7.1)
$$
where $\Omega_K$ is the Keplerian angular velocity.  Their argument
was that viscosity depends on two quantities: (i) the speed $\s$ of
the turbulent eddies, which is likely to be limited by the
sound speed $c_s$, and (ii) the mean free path $l$ which is likely to
be no larger than the vertical scale height of the disk, the latter
quantity being $\sim c_s/\Omega_K$.  Since $\nu \sim \s l$, they thus obtained
equation (3.7.1) with $\alpha$ limited to be $\sles\ 1$.

In \S3.6 we have developed a fairly complete treatment of viscous
interactions between particles in a differentially rotating system and
we can therefore quantify equation (3.7.1) more carefully.  Our theory has
two parameters.  One of these is the {\it r.m.s.} particle velocity $\s$.  In
the context of disk viscosity, our ``particles'' are turbulent eddies or
blobs, and in a purely hydrodynamic situation we expect the blob
velocity $\s$ to be smaller than the local sound speed,
otherwise it would be dissipated by shocks. Let us therefore write
$$
\s = \alpha_1 c_s, \q \alpha_1 \leq 1.\eqno(3.7.2)
$$
Our second parameter is the mean free time $\t$ between collisions.  Since
we cannot a priori assign any limits to $\t$,
let us simply express it in terms of the rotation frequency $\Omega$,
$$
\t = {\alpha_2 \o \Omega}, \eqno(3.7.3)
$$
where we do not restrict $\alpha_2$ to any particular range.  We showed
in \S 3.6 that the viscosity depends on two
frequencies, $A$ and $\k$, which are given by
$$
{2A\o \Omega} = {R\o \Omega} {d\Omega\o dR} = {d\ln\Omega \o d\ln
R},\eqno(3.7.4)
$$
$$
{\k^2\o \Omega^2} = {2\o \Omega R} {d\o dR}(\Omega R^2) = 2 {d\ln(\Omega R^2)
\o d\ln R}.\eqno(3.7.5)
$$

Substituting these relations into equation (3.6.13) we find the following
result for the coefficient of viscosity
$$
\nu = {\alpha ^2_1 \alpha_2\o 3 + 2\alpha^2_2(d\ln\Omega/d\ln R)^2 + 24
\alpha^2_2(d\ln\Omega R^2/d\ln R)} {c^2_s\o \Omega} \equiv \alpha
{c^2_s\o \Omega}, \eqno(3.7.6)
$$
which gives an expression for the Shakura-Sunyaev $\alpha$ parameter
in terms of $\a_1$ and $\a_2$.
In the particular case of a Keplerian disk, where $\Omega (R)\propto
R^{-3/2}$,
we have $2A/\Omega = -3/2$ and $\k^2/\Omega^2 = 1$, and (3.7.6) becomes
$$
\nu_K = {\alpha^2_1 \alpha_2 \o 3 + 33\alpha^2_2/2}~ {c^2_s \o \Omega}
\equiv \alpha_K {c^2_s\o \Omega}.\eqno(3.7.7)
$$
The most
interesting feature of these relations is the fact that $\nu$ and $\nu_K$
each has an absolute maximum value regardless of the value of $\alpha_2$.
Equation (3.7.7) for instance reaches its maximum value when
$\Omega\tau\equiv \alpha_2 = \sqrt{2/11}$ at which point we have
$$
(\alpha_K)_{max} = {\alpha_1^2\o\sqrt{198}}=0.071\alpha^2_1.
\eqno(3.7.8)
$$
This is an extremely small limit for $\alpha_K$ considering that $\alpha_1$
is itself limited to lie below unity. Note that if $\alpha$ is
defined alternatively through the stress tensor relation
$\ov{u_1u_2}=-\alpha c_s^2$, then the above bound should
be multiplied by $3/2$.

To demonstrate how small $\ak$ is in a Keplerian disk we show in
figure 9 the variation of $\ak$ as a function of the parameter
$\alpha_2\equiv\Omega\tau$.  We see that, if $\a_2$ lies more than a
factor of few outside the optimum value of $\sqrt{2/11}$, then $\ak$
falls even further.  Also, $\a_1$ itself is presumably somewhat less
than unity since turbulent blobs are likely to have a range of speeds
with the fastest blobs limited to speeds less than $c_s$.  Thus, we
conclude that turbulent viscosity in accretion disks, when expressed
in the Shakura-Sunyaev form (3.7.1), is likely to have values of
$\ak$ which are somewhat smaller than $0.07$.  Lower values of $\ak$
imply disks with larger surface densities and spectra that are more
nearly thermal.

In regions away from the Keplerian region of a disk, for example in
the inner boundary layer, we need to use the full expression given in
equation (3.7.6).  We see that in regions of large shear the
effective $\a$ will be much smaller than in the Keplerian case.
Figure 9 shows a case where $d\ln\Omega/d \ln R=5$, a modest value by
the standards of boundary layers (e.g.  Narayan and Popham 1993,
Popham et al. 1993).  Note how small the viscosity coefficient
becomes in this case.  When we add to this the fact that the boundary
layer region has radial flows which also reduce $\nu$ through the
causality factor discussed in \S 3.4, we see that accretion disk
boundary layers will in general have very weak viscosity.  The effect
that this will have on the structure of the boundary layer is yet to
be investigated.  The role of third moments, which may not be
negligible in boundary layers, is also unclear.

The formulae we have developed are based on the assumption of a
constant mean free time, and one may wonder whether the results would
change significantly with other scattering laws.  However, to leading
order, different scattering laws should all be describable through an
effective mean free time, and since we did not restrict $\Omega\tau$ in
any way our limit on $\a$ would appear to be robust.  The point is
that, ultimately, the limit on $\a$ comes from the stress-limiter,
which inhibits the viscous stress from exceeding the pressure, coupled
with geometrical factors which further limit the shear stress because
of the epicyclic motion.  These are sufficiently general principles
that we cannot envisage violating the limits shown in figure 9 by any
significant factor.  Obviously, if the shear stress is produced by
non-hydrodynamic effects, e.g.  magnetic fields, then  our
limit could be exceeded.

\vfill\eject
\centerline{\bf 4. SUMMARY AND DISCUSSION}
\bigskip

In this work we have attempted to incorporate causality into the
theory of transport phenomena.  We have developed equations for
particle diffusion in one and three dimensions which are an
improvement over the standard diffusion equation whenever gradients
are large and causality limits the flux. We have also generalized the
equations of viscous hydrodynamics to include situations where
velocity gradients are large or time variations are rapid. In such
cases the standard Navier-Stokes equation breaks down.

Our approach is based on taking moments of the Boltzmann equation and
using an approximate form for the collision term (equation 2.1.3).  We
assume that particles collide with a mean free time $\t$ which is
independent of velocity. Furthermore, we assume that each scattering
completely isotropizes the velocity distribution of the particles.
These assumptions permit us to write simple expressions for the
velocity moments of the post-scattering distribution function $f_0$ in
terms of the moments of the pre-scattering function $f$. The theory is
constructed to handle situations where $f$ differs appreciably from
$f_0$.  This is an important departure from the usual perturbative
method of analyzing transport phenomena where the quantity $(f-f_0)$
in equation (2.1.3) is assumed to be small compared to the equilibrium
distribution $f_0$.  The non-perturbative nature of our theory allows
us to write down moments of the Boltzmann equation which are valid
even in situations that involve large departures from equilibrium.

The first example we have presented is the case of light particles
diffusing in one dimension in a fixed background (\S2.1, \S2.2).  We
take the zeroth and first moments of the Boltzmann equation and close
the equations by setting the second moment of the velocity, $\vbb$,
to a constant.  This leads to a modified diffusion equation (2.1.11)
which differs from the usual diffusion equation in having a wave
operator instead of the standard Laplacian operator.  The presence of
the second time derivative in the wave operator enforces causality
(Israel and Stewart 1980, Schweizer 1984).  This term arises from a
time derivative term in the first moment equation, which is neglected
in deriving the standard diffusion eqution.  We therefore deduce
that, if we wish to preserve causality, it is important to retain
consistently all time derivatives of higher moments.  We follow this
prescription in the various problems we have analyzed and we find
that we obtain equations that are causal in all cases.

The Green's function of the one dimensional diffusion problem reveals
several interesting features.  First, it vanishes outside the
particle ``velocity cone,'' $|x|>\s t$, where $x$ is distance from
the point of injection of the particles and $\s$ is the {\it r.m.s.}
velocity of the particles.  This explicitly demonstrates the causal
nature of the equations.  Secondly, the Green's function has
$\delta$-functions exactly at the two edges of the velocity cone.
This shows that the theory effectively replaces the full velocity
space distribution $f(v)$ by two $\delta$-functions at $v=\pm\s$.
This simplification means that the theory cannot treat phenomena like
phase mixing which arise from a continuous distribution of
velocities.  If such effects are important in any application then
the theory we have presented is inadequate.  One should either
integrate the single velocity Green's function over the particle
distribution function or use higher order moments (see equations
[2.1.19] \& [2.1.20]).  The theory also ignores the tail of high
velocity particles which are normally found in a real distribution
function $f(v)$. Real gases do not diffuse with an exact cutoff in
space. While the bulk of the signal is limited to a speed equal to
the {\it r.m.s.} speed $\s$ of the particles, there is always an
exponentially small number of particles at large velocities
propagating faster than $\s$.  By eliminating the high velocity tail
altogether our theory ignores this weak precursor signal and
therefore has an exact cutoff of signals at a maximum propagation
speed $\s$.

Leaving aside the issues of phase mixing and the exponential tail, we
show that the theory does well in describing the behavior of the bulk
of the particles.  In fact, when all particles have a constant mean
free time $\tau$ the theory gives the mean square distance $\lx$
traversed by particles {\it exactly}, regardless of their velocity
distribution function. The agreement is perfect both at early times,
when the particles are streaming freely, and at late times, when they
have undergone many scatterings. Furthermore, even for a more general
scattering law we can usually find an equivalent $\tau$ such that our
theory provides a satisfactory description of the true evolution of
$\lx$ (see figure 2).

It is a straightforward matter to generalize the theory to diffusion
in three dimensions. The only complication originates from the fact
that the minimal extension of the one-dimensional closure relation
requires a somewhat more extreme assumption in three dimensions,
namely that the velocity second moment tensor is not only constant but
also isotropic. This assumption leads to a causal but somewhat
unphysical Green's function with negative densities.  Despite
this unsatisfactory feature, the theory provides an excellent
description of the evolution of the mean square distance $\lr$
travelled by the particles.

An interesting special case of the three dimensional problem is when
the density gradient is limited to a single direction, say the $x_1$
axis.  In this case, we find that our three-dimensional diffusion
equation reduces to a form that is identical to the one dimensional
equation except that the speed of propagation of signals is no longer
the {\it r.m.s.} particle speed $\s$ but rather $\s/\sqrt{3}$.  In
this particular case, it is clear that we effectively replace the
full distribution function $f(v_1, v_2,v_3)$ by one that is
restricted to two planes in velocity space, i.e.
$v_1=\pm\s/\sqrt{3}$.  We have already seen the limitations
associated with such an approximation, namely the inability to model
phase mixing and signal propagation by the exponential tail of fast
particles.  But apart from this, we expect the one-dimensional
projection of the three-dimensional equation to behave physically and
to predict the behavior of the majority of particles in a
satisfactory way.  In particular, the Green's function for this
problem does not have unphysical negative densities.

In \S3 we have used our method to derive a general causal equation
(3.1.11) for the evolution of the stress tensor. Since the stress
tensor is of second order in the velocities, we take the zeroth,
first and second moments of the Boltzmann equation, assuming as in
the diffusion problem a constant mean free time $\t$.  We also set
the third moments of the velocity to zero, which allows us to close
the moment sequence and to derive a self-contained set of equations.
The main additional feature in our model of the scattering is the
introduction of a ``cooling'' factor $\xi$ (see equation 3.1.4)
whereby scattering events are allowed to be inelastic.  This
modification allows us to model cooling processes where some of the
kinetic energy of colliding particles is converted into another form
of energy, e.g.  radiation, and removed from the system. By including
cooling, we are able to investigate steady state problems even when
there is heating due to stresses.

Before proceeding to describe our results on the stress tensor, we
discuss the effect of the various approximations we have made.  The
physical systems we have investigated involve velocity gradients
along a single direction, a case where we can expect the theory to be
well-behaved since we know that the particle diffusion problem admits
a physical solution.  The theory assumes a constant mean free time
for all particles and therefore involves a simplification of the real
collision integral.  However, as we have shown in detail in the
diffusion problem, and with some limited numerical experiments in the
stress problem as well (figure 6), general scattering laws can
usually be reduced to an effective constant $\tau$ with very little
qualitative differences.  Therefore, the assumption of a constant
$\t$ is unlikely to be a serious limitation.

The most serious simplification of our theory is the neglect of third
moments.  This approximation is reasonable when the velocity
distribution has a reflection symmetry, or when the mean velocity
profile is nearly linear in the vicinity of the region of interest.
As we show in the Appendix, the third moment terms will be small
provided the dimensionless second velocity derivative,
$\s\t^2\p^2\overline{v}/\p x^2$, is small compared to either unity or
the square of the dimensionless shear (equation A10).  There is
another similar condition on the third derivative of the velocity
(equation A11).  These conditions may not be satisfied in highly
transient flows or in the vicinity of shocks, but should be valid in
smooth steady state flows and especially in rotating systems where the
distortion of the particle velocity distribution tends to be limited
(see figure 8).  It is important to emphasize that the neglect of the
third moments does {\it not} require the velocity gradient to be
small; the theory we have developed is valid for arbitrarily large
gradients as long as the gradient itself is spatially constant, i.e.
the second derivative of the velocity is sufficiently small.

Subject to the above caveats, equations (3.1.6) \& (3.1.11) may serve
as potential candidates to replace the Navier-Stokes equation whenever
spatial or temporal derivatives are large. We have not tested these
equations in their full generality.  Rather, we have concentrated on
special cases involving steady state shear flows to demonstrate
several physical effects involving causality in viscous transport
processes.

In \S 3.2--3.4, we derive some general results for the shear stress
in a steady shear flow.  According to the standard theory, a velocity
shear $\p\ov{v_2}/\p x_1\equiv 2A$ gives rise to a shear stress $n\nu
(2A)$, where $n$ is the density and $\nu$ is the kinematic
coefficient of viscosity.  Our causal theory reproduces this result
in limit that $A$ is small, with $\nu=\s^2\t/3$, but more generally
reveals three distinct new effects described below:

\item{1.} In \S 3.2 we consider steady shear of arbitrary magnitude $2A$
without advection and show that the shear stress is given by
$n\nu(2A)(1+2[2A\tau]^2/3)^{-1}$ (equation 3.2.10).  This expression
gives a significantly smaller stress than the standard result
mentioned above when the dimensionless shear $2A\t$ is large.  Whereas
the standard relation for the shear stress diverges in the limit of
large shear, the modified expression reaches a maximum value, which is
less than the pressure, at a finite value of the dimensionless shear
$2A\t$ and reduces for larger values of the shear.  This behavior is a
consequence of causality.  The particles at any given point typically
have originated a mean free path away.  Therefore, for a large shear,
they have a large transverse velocity spread
$\ov{u_2^2}\sim\sigma^2\gg\ov{u_1^2}$.  This translates to a stress
limit $\vert\ov{u_1u_2}\vert< (\ov{u_1^2}\ \ov{u_2^2})^{1/2}
\ll\sigma^2$.  We have tested our result for the shear stress through
numerical particle simulations spanning a range of values of the
dimensionless shear $2A\t$ (\S 3.3).  The agreement is essentially
perfect (to within numerical precision) when all particles have the
same mean free time.  Even when we instead consider the mean free path
to be constant, our theory agrees well with the numerical simulations
(figure 6).  In particular, the theory predicts the maximum value of
the shear stress accurately.  Note that in the case of the shear
stress, not only does causality set a limit to the maximum stress, it
actually causes a reduction of the stress in the limit of large shear.
This asymptotic behavior is very different from the behavior of scalar
particle diffusion (e.g.  Levermore and Pomraning 1981, Narayan 1992),
where the flux asymptotically approaches to a constant value in the
limit of a large density gradient.

\item{2.} When a constant advection velocity, $\ov{v_1}$, is added to the
steady shear flow, we find that the shear stress is reduced further by
a second suppression factor $(1-3\ov{v_1}^2/\ov{u_1^2})$ (cf. eq.
[3.4.4]).  When there is advection, the shear stress has to be
communicated by particles that move upstream faster than the advection
speed $\ov{v_1}$.  The flux of such upstream moving particles reduces
as the advection speed increases and cuts off when $\ov{v_1}$ exceeds
the {\it r.m.s.} velocity ${(\ov{u_1^2})^{1/2}}$ of particles in the
direction of the flow.  A similar qualitative result was obtained by
Narayan (1992) who showed that causality leads to a reduction and
ultimate cutoff of viscosity when there is advection.  Narayan
considered somewhat general velocity distribution functions whereas
our discussion corresponds to a very simple velocity distribution
consisting of two $\delta-$functions at $u_1=\pm(\uxxb)^{1/2}$.  On
the other hand, Narayan assumed that the shear is small, whereas we
allow arbitrarily large velocity gradients.  One other difference is
that Narayan's expression for the viscosity coefficient left
undetermined the critical advection velocity at which the viscosity
will vanish.  The present theory shows that, at this level of
approximation, the cutoff occurs exactly at the sound speed
$\s/\sqrt{3}$.  We caution, however, that in real gases the advection
factor may not involve a sudden truncation of the viscosity
coefficient but rather an exponential cut off as $\ov{v_1}$ exceeds
$(\uxxb)^{1/2}$.  This is because of the high velocity tail in the
particle distribution function which may be able to transmit a weak
signal upstream even at large advection speeds.

\item{3.} We have also investigated the effect of a non-uniform advection and
find that the shear stress is modified by a third factor,
$(1+2\tau\partial\ov{v_1}/\partial x_1)^{-1}$~~(cf. eq.  [3.4.4]).
This shows that the shear stress is suppressed when the flow expands,
i.e. when $\partial{\ov{v_1}}/\partial x_1>0$, but is enhanced in the
presence of compression.  Curiously, the stress diverges when
$2\tau\partial{\ov{v_1}}/\partial x_1=-1$, but this is not a serious
problem since it refers to an unphysical limit where the gas is
compressed to an infinite density in a collision time.  The same
limiting compression appears also in the context of our analysis of
bulk viscosity in \S 3.5.  Kato and Inagaki (1993) have independently
noted that expansion or contraction of the advecting flow can modify
the shear stress.

In \S 3.6, we extend the analysis to rotating flows and find that the
above results for the shear stress are further modified.  We find that
the shear stress in a rotating flow in the absence of advection is
given by $n\nu(2A)(1+2[2A\tau]^2/3+4[\kappa\tau]^2)^{-1}$, where
$\Omega$ is the angular velocity and $\kappa^2=4\Omega(A+\Omega)$ is
the square of the epicyclic frequency.  Thus, in the presence of
rotation, the shear stress is reduced compared to the classical
diffusion formula $n\nu(2A)$ when either the shear or the epicyclic
frequency is large compared to the collision frequency $1/\t$. The
suppression effect due to $\kappa$ has been noted in previous studies
by Goldreich and Tremaine (1978) and Kato and Inagaki (1993).  Kato
and Inagaki's approach is very similar to ours, whereas Goldreich and
Tremaine treat two particle inelastic collisions in much greater
detail, but they close the moment equations by assuming that the
distribution function in velocity space has a triaxial Gaussian shape.
When $\kappa\tau\gg1$, a typical particle undergoes many epicycles
within a collision time and consequently moves radially only across
the limited scale of an epicycle.  The effective mean free path is
therefore smaller than in the non-rotating case by a factor $\sim
\kappa\tau$, and the stress is reduced by the square of this factor.
For the same reason the distribution of particle velocities in phase
space does not stretch indefinitely as in the non-rotating case
(figures 5 and 7), but rather reaches a limiting oval shape as
$\Omega$ increases beyond $1/\tau$ (figure 8).  The new feature that
comes out of our theory is that the viscosity coefficient is
suppressed not just by the dimensionless epicyclic frequency
$\kappa\t$ but by a combination of the dimensionless shear $2A\t$ and
$\kappa\t$.  This result is of potential importance in astrophysics.
For instance, in Keplerian disks, the two frequencies $2A$ and
$\kappa$ are of comparable magnitudes.

Finally, in \S 3.7, we have applied our results to thin accretion
disks, where we find that the causal limit on the shear stress leads
to a constraint on the value of the disk viscosity coefficient $\nu$.
This limit can be translated to an upper bound on the dimensionless
$\alpha$-parameter defined by $\nu=\alpha c_s^2/\Omega$, where $c_s$
is the sound speed in the disk.  For hydrodynamic turbulence in a
Keplerian disk we find that $\alpha<\alpha_1^2/{\sqrt{198}}<0.07$ for
any value of $\Omega\tau$ (see figure 9), where $\alpha_1<1$ is the
ratio between the {\it r.m.s.} turbulent blob speed and the sound
speed in the disk.  This bound can be exceeded only if the turbulence
is non-hydrodynamic (e.g. magnetic) and $\alpha_1>1$. Note that our
upper bound should be scaled up by a factor of 3/2 if $\alpha$ is
instead defined through the stress tensor relation
${\ov{u_1u_2}}=-\alpha c_s^2$.  The upper limit on $\alpha$ is much
stronger in boundary layers where the velocity shear is larger than
Keplerian.

To conclude, we note three issues related to viscous flows that we
have not addressed in this paper, namely (i) the effect of advection
in rotating flows (where the physics appears to be a little different
than in the non-rotating case because of the Coriolis acceleration),
(ii) the structure of viscous shocks, and (iii) time-dependent
effects.  However, the basic equations of our theory, especially the
stress-evolution equation (2.1.11), appear to be capable of dealing
with these issues.  Work along these directions may provide further
insight into the role of causality in astrophysical accretion disks.

\bigskip
\bigskip
\bigskip
\bigskip
\bigskip
\noindent
{\bf Acknowledgments:} RN is grateful to Jeremy Goodman for
fascinating discussions several years ago on the physics of diffusion,
and to Scott Grossman for valuable discussions on the Boltzmann
equation.  AL thanks Russell Kulsrud for insightful comments.  We are
grateful to S. Kato and S. Inagaki for sending a preprint of their
paper ahead of publication.  This work was supported in part by grant
AST 9148279 from the NSF.

\np
\centerline{\bf Appendix}
\bigskip

In the following we rederive the results given in \S 3.2 for the
components of the stress tensor in the presence of an arbitrary
velocity shear, $\p\ov{v_2}/\p x_1\equiv 2A$.  We employ here a
different, and possibly more transparent, approach in which we follow
the trajectories of individual particles.  We then extend this
approach to investigate under which conditions our assumption of
vanishing third moments is valid.  This helps us identify the
limit of validity of the theory developed in the paper.

Consider the particles at a particular plane, $x_1=0$, at a given
time and label the particles by the coordinate $x_1$ at which they
last suffered a scattering.  Let the relative velocity component of a
particle immediately after the last scattering be given by $(u_1)_0$,
$(u_2)_0$, $(u_3)_0$.  The relative velocity components at position
$x_1=0$ are then
$$
u_1=(u_1)_0,\qquad u_2=(u_2)_0+2Ax_1+{1\o 2}\vsec
x_1^2\cdots,\qquad u_3=(u_3)_0,\eqno (A1)
$$
where we have Taylor-expanded the mean velocity $\ov{v_2}$ as a function
of $x_1$.  By the assumption that the scattering is isotropic,
we have
$$
\ov{(u_1^2)_0}=\ov{(u_2^2)_0}=\ov{(u_3^2)_0}.\eqno (A2)
$$
Furthermore, since $u_1$ and $u_3$ of a particle do not vary
in between scatterings, we have
$$
\ov{u_1^2}=\ov{(u_1^2)_0},\qquad\ov{u_3^2}=\ov{(u_3^2)_0}.\eqno (A3)
$$

Consider a particle at $x_1=0$ with a particular value of $u_1$.
Given our assumptions of constant density, constant second moments, and
constant mean free time, the
probability that this particle had its last scattering
between $x_1$ and $x_1+dx_1$ is given by
$$
p(x_1)dx_1={1\o |u_1|\t}\exp (x_1/u_1\t)dx_1, ~~x_1u_1<0.\eqno (A4)
$$
We then see that, averaged over all particles at $x_1=0$,
odd moments of $x_1$
such as $\langle x_1\rangle$ and $\langle x_1^3\rangle$ vanish, while
even moments become
$$
\langle x_1^2\rangle =2\ov{u_1^2}\t^2,\qquad
\langle x_1^4\rangle =24\ov{u_1^4}\t^4,\qquad\cdots\eqno (A5)
$$

Let us consider now the case discussed in \S 3.2, where the velocity
profile is exactly linear with a slope $2A$ and where the second
derivative of the velocity is zero.  From equation (A1), we find the
second moment of $u_2$ to be given by
$$
\ov{u_2^2}=\ov{(u_2^2)_0}+4A^2\langle x_1^2\rangle
=(1+8A^2\t^2)\ov{u_1^2}.\eqno (A6)
$$
Using now the relation, $\s^2=\ov{u_1^2}+\ov{u_2^2}+\ov{u_3^2}$,
we find that
$$
\ov{u_1^2}=\ov{u_3^2}={\s^2\o3+8A^2\t^2},\qquad
\ov{u_2^2}={1+8A^2\t^2\o3+8A^2\t^2}\s^2.\eqno (A6)
$$
Similarly, we find the velocity moment $\ov{u_1u_2}$ to be given by
$$
\ov{u_1u_2}=\ov{(u_1u_2)_0}+\langle 2Ax_1(u_1)_0\rangle
=-2A\t\ov{u_1^2}=-{2A\t\o3+8A^2\t^2}\s^2.\eqno (A7)
$$
These relations are identical to the results derived in \S 3.2.

Let us proceed next to a case where the second velocity derivative
$\p^2\ov{v_2}/\p x_1^2$ and higher derivatives are
not zero but small and let us estimate the
contribution that they make to the shear-stress.
This is best accomplished by considering the $u_1$--$u_2$ component of
equation (3.1.11) which is given below:
$$ \ov{u_1u_2} = -2\tau A \ov{u_1^2} - \tau{\partial\ov{u_1^2u_2}
  \over\partial x_1}. \eqno(A8)$$
We can estimate $\ov{u_1^2u_2}$ using equations (A1) and (A4).
Substituting this into equation (A8) we find that the modification
to the shear stress is small if the following
condition on the derivatives of $\ov{v_2}$ is satisfied:
$$\left| {\partial\over\partial x_1}\left[ \tau^2\ov{u_1^4}
\vsec\right]\right| \ll 2A\ov{u_1^2}, \eqno(A9) $$
Using the Gaussian closure relation, $\ov{u_1^4}=3(\ov{u_1^2})^2$,
this means that we require
$$ \s\t^2\left|\vsec\right|\ll {(3+8A^2\t^2)\over\sqrt{24}}, \eqno (A10) $$
and
$$\s^2\t^3\left|{\partial^3 \ov{v_2}\over\partial x_1^3}\right| \ll
  {2A\t(3+8A^2\t^2)\over 3}. \eqno(A11) $$

We thus see that, when the dimensionless shear $2A\t\ll1$, the
condition for the third moments to have a negligible effect is that
(i) the dimensionless second velocity derivative in the left-hand side
of equation (A10) must be small compared to unity, and (ii) the
dimensionless third derivative in (A11) must be small compared to
$2A\t$.  When $2A\t\gg1$, both of these criteria are weakened by a
factor of $A^2\t^2$.

For a rotating system the value of $x_1$ is limited by the radial
extent of an epicycle $\sim u_1/\kappa$.  In the case of a thin disk
with a scale-height $h\ll r$ and a Keplerian velocity profile, the
conditions (A10) and (A11) will be automatically satisfied since the
left hand side is always of order $(h/r)\ll1$.  Thus the limit on
$\alpha$ which was derived in \S 3.7 under the assumption that the
third velocity moments vanish is self-consistent.

\np
\centerline{\bf REFERENCES}

\medskip
\noindent
Binney, J., \& Tremaine, S. 1987, Galactic Dynamics,
(Princeton: Princeton Univ. Press), pp. 269-271
\medskip
\noindent
Goldreich, P., \& Lynden-Bell, D. 1965, MNRAS, 130, 125
\medskip
\noindent
Goldreich, P., \& Tremaine, S. 1978, Icarus, 34, 227
\medskip
\noindent
Grossman, S. A., Narayan, R., \& Arnett, D. 1993, ApJ, 407, 284
\medskip\noindent
Israel, W., \& Stewart, J. M. 1980, in General Relativity and Gravitation,
vol. 2, ed. A. Held (New York: Plenum)
\medskip
\noindent
Kato, S. 1970, Publ. Astron. Soc. Japan, 22, 285
\medskip
\noindent
Kato, S., \& Inagaki, S. 1993, Publ. Astron. Soc. Japan, submitted
\medskip
\noindent
Levermore, C. D., \& Pomraning, G. C. 1981, ApJ, 248, 321
\medskip
\noindent
Max, C. E. 1981, Physics of Laser Fusion, LLNL report \# UCRL-53107
\medskip
\noindent
Morse, P. M., \& Feshbach, H. 1953, Methods of Theoretical Physics,
Part I, (New-York: McGraw-Hill), \S 7.4, pp. 857-869
\medskip
\noindent
Nagel, W., \& M\'esz\'aros, P. 1985, J. Quant. Spectrosc. Radiat. Transfer,
34, 493
\medskip
\noindent
Narayan, R. 1992, ApJ, 394, 261
\medskip
\noindent
Narayan, R., \& Popham, R. 1993, Nature, 362, 820
\medskip
\noindent
Narayan, R., Goldreich, P., \& Goodman, J. 1987, MNRAS, 225, 695
\medskip
\noindent
Popham, R., \& Narayan, R. 1992, ApJ, 394, 255
\medskip
\noindent
Popham, R., Narayan, R., Hartmann, L., \& Kenyon, S. 1993, ApJ, 415, L127
\medskip
\noindent
Reif, F. 1965, Fundamentals of Statistical and Thermal Physics,
(New-York: McGraw-Hill), pp. 560-567
\medskip\noindent
Schweizer, M. A. 1984, J. Phys. A: Math. Gen., 17, 2859
\medskip
\noindent
Shakura, N. I., \& Sunyaev, R. A. 1973, A\&A, 24, 337
\medskip
\noindent
Shu, F. H. 1992, The Physics of Astrophysics, Volume II: Gas Dynamics,
(Mill Valley: University Science Books), pp. 33-34
\medskip
\noindent
Syer, D., \& Narayan, R. 1993, MNRAS, 262, 749

\np
\noindent{\bf FIGURE CAPTIONS}

\bigskip
\noindent
{\bf Fig. 1:} Comparison between the Green's function for the
one-dimensional diffusion of a Maxwellian distribution function
(dot-dashed line), the Green's function of the standard diffusion
equation (dotted), and the Green's function (2.1.13) of the modified
diffusion equation (2.1.11) (solid).  The curves are normalized to a
unit area and are shown at two different times: $t=0.1\tau$ (upper
panel) and $t=10\tau$ (lower panel), where $\tau={\rm const}$ is the
collision mean free time. In the lower panel, the two
$\delta-$functions which propagate at $v=\pm \s$ have not been
plotted since they are suppressed by $e^{-t/2\tau}=e^{-5}$.

\bigskip
\noindent
{\bf Fig. 2:} The evolution of the {\it r.m.s.} particle distance
$\langle x^2\rangle^{1/2}$ for one-dimensional diffusion away from an
instantaneous point source.  Length and time are normalized by the
mean free path $l$ and the mean free time $l/\sigma$, respectively.
The different curves compare the exact result for a one-dimensional
Maxwellian distribution of particles with a constant mean free path
(solid line) to our one-dimensional causal model that assumes a
constant mean free time $\tau=(2/\pi)^{1/2}l/\s$ (short dashed) and
the standard diffusion approximation result (long dashed).  The
causal model fits the true behavior very well.  Note that our model
describes the evolution of $\langle x^2\rangle$ precisely for any
system of particles that has a constant mean free time.

\bigskip
\noindent
{\bf Fig. 3:} The same as in figure 2 but for the {\it r.m.s.}
particle distance $\langle r^2\rangle$ in three-dimensional
diffusion.

\bigskip
\noindent
{\bf Fig. 4:} Evolution of the second velocity moments with time. The
results were obtained from a numerical simulation of a linear shear
flow with $10^5$ particles, corresponding to $2A=1$,
$(1-\xi)\sigma^2=1$, and $\tau=1$. Note that the moments achieve
their steady-state values $\ov{u_1^2}=\ov{u_3^2}=1/3$, $\ov{u_2^2}=1$
and $\ov{u_1u_2}=-1/3$ in less than 10 mean free times.

\bigskip
\noindent
{\bf Fig. 5:} Phase-space distribution of particles in the
$u_1$-$u_2$ plane for steady-state flow with a linear shear.  Results
are shown for three different shear amplitudes $2A\tau$, namely: (a)
0.1, (b) 1.0, and (c) 10, where $\tau$ is the mean free time, taken
to be independent of particle velocity.  Note the strong distortion
of the velocity distribution at large shear.

\bigskip
\noindent
{\bf Fig. 6:} The stress in a steady-state shear flow of particles
with a constant mean free path $l$. The points show
$\ov{u_1u_2}/\sigma^2$ as a function of the shear amplitude
$2Al/\sigma$, according to numerical simulations with $10^3$
particles each. The curves show the prediction of equation (3.2.10) when
we set $\tau=l/\sigma$ (dashed line) or $\tau=0.55l/\sigma$ (solid
line).  The maximum stress is predicted accurately by our model even
though it assumes a constant mean free time $\tau$ instead of a
constant $l$.

\bigskip
\noindent
{\bf Fig. 7:} Phase-space distribution of particles in the
$u_1$-$u_2$ plane for steady-state flow of particles with a linear
shear $2A$ and a constant mean free path $l$. The different panels
correspond to different shear amplitudes $2Al/\sigma$, namely: (a)
0.1, (b) 1.0, and (c) 10.

\bigskip
\noindent
{\bf Fig. 8:} Phase space distribution of particles in the
$u_1$-$u_2$ plane for a steady shear in a rotating Keplerian disk.
Results are shown for four values of $\Omega\tau$, namely: (a) 0.1,
(b) 0.4264, (c) 2.0, and (d) 10.  Case (b) yields the maximum value
of $\ov{u_1u_2}$. Note the limited distortion of the velocity
distribution for large shear, in contrast to the cases shown in
figures (5) and (7).  This is because the asymptotic form of the
velocity distribution in a rotating flow under a large shear is
moderated by the epicyclic motion (see equations [3.6.10]-[3.6.12]).

\bigskip
\noindent
{\bf Fig. 9:} The value of the $\alpha$-viscosity parameter for
hydrodynamic turbulence in thin disks (cf. eq. [3.7.6]) as a function
of $\alpha_2\equiv \Omega\tau$, where $\Omega$ is the disk angular
velocity and $\tau$ is the mean time between collisions of turbulent
blobs. The value of $\alpha$ is divided by $\alpha_1^2$, where
$\alpha_1\equiv \sigma/c_s<1$ is the ratio between the {\it r.m.s.}
blob velocity and the sound speed in the disk. Results are presented
for two rotation profiles: a Keplerian disk (solid line) and a
``boundary layer'' profile with $2A=5\Omega$ (dashed).
\bye